\documentclass[prd,nofootinbib,11pt]{revtex4}
\usepackage[colorlinks=true,linkcolor=red,citecolor=blue]{hyperref}
\usepackage{amsmath}
\usepackage{slashed}
\usepackage{multirow}
\usepackage{array}
\usepackage{adjustbox}
\usepackage{bm}
\usepackage[utf8]{inputenc}
\usepackage{enumitem}
\usepackage{makecell}
\renewcommand{\arraystretch}{1.3}
\begin{document}
\title{\boldmath New Physics in $\bm{b\to s\ell\ell}$ anomalies and its implications for the complementary neutral current decays}
\author{Faisal Munir Bhutta$^1$}
\email{faisalmunir@bjut.edu.cn}
\author{Zhuo-Ran Huang$^{2,3}$}
\email{zhuoran.huang@apctp.org}
\author{Cai-Dian L\"u$^{4,5}$}
\email{lucd@ihep.ac.cn}
\author{M. Ali Paracha$^6$}
\email{aliparacha@sns.nust.edu.pk}
\author{Wenyu Wang$^1$}
\email{wywang@bjut.edu.cn}
\affiliation{$^1$Institute of Theoretical Physics, Faculty of Science, Beijing University of Technology, Beijing 100124, China\\
$^2$Université Paris-Saclay, CNRS/IN2P3, IJCLab, 91405 Orsay, France\\
$^3$Asia Pacific Center for Theoretical Physics, Pohang, 37673, Korea\\
$^4$Institute of High Energy Physics, Chinese Academy of Sciences, Beijing 100049, China\\
$^5$School of Physics, University of Chinese Academy of Sciences, Beijing 100049, China\\
$^6$Department of Physics, School of Natural Sciences (SNS), National University of Sciences and Technology (NUST), Sector H-12 Islamabad, Pakistan}

\begin{abstract}
We study the Standard Model and the new physics predictions for the lepton-flavour-universality violating (LFUV) ratios in various
$b\to s \ell^+\ell^-$ channels with scalar, pseudoscalar, vector, axial-vector, and $\Lambda$
baryon final states, considering both unpolarized and polarized final state hadrons.
In order to formulate physical observables, we use the model independent effective Hamiltonian approach
and employ the helicity formalism. We provide the explicit expressions of the helicity amplitudes in terms of the Wilson coefficients
and the hadronic form factors by using the same kinematical configuration and polarization conventions for all the decay channels.
We perform the numerical analysis with new physics scenarios selected from the recent global fits to $b\to s\ell^+\ell^-$ data, having specific new physics model interpretations. We find that some of the LFUV ratios for these complementary channels in different kinematical regions
have high sensitivity to new physics and the future measurements of them in Belle II and LHCb experiments, along with testing new physics/LFUV,
can help to distinguish among some of the different new physics possibilities.
\end{abstract}

\maketitle

\section{Introduction}\label{intro}
Flavour-changing neutral-current (FCNC) processes involving $b\to s \ell^+\ell^-$ quark level transitions can play
a pivotal role in the indirect searches of physics beyond the Standard Model (SM). These transitions are CKM and
loop suppressed within the SM and therefore have high sensitivity to potential new physics (NP) effects.
Interestingly, recent experimental data on neutral current decays induced by $b\to s \ell^+\ell^-$
transitions have pointed towards several observables in tension with the SM predictions. Due to this fact, these
transitions currently stand among the most promising indications of NP.

%and because of the reported anomalies stand among the most promising indications of NP.
%An indirect hunt of new physics (NP) is underway at the dedicated flavour experiments such as LHCb and Belle.

%The reported observables can be grouped into two sets: $b\to s \mu^+\mu^-$ observables that include only muons,
%called as lepton-flavour dependent (LFD) observables, and the other known as lepton-flavour-universality violating
%(LFUV) observables that involve both muons and electrons. The set of LFD observables contains several angular
%observables including the so-called $P_5^{\prime}$ anomaly \cite{Descotes-Genon:2013vna,Descotes-Genon:2013wba}, and
%the differential branching fractions of the $B\to K\mu^+\mu^-$ \cite{Aaij:2014pli}, $B\to K^{\ast}\mu^+\mu^-$ \cite{Aaij:2014pli,Aaij:2013iag,Aaij:2016flj}, %and $B_s\to \phi\mu^+\mu^-$ \cite{Aaij:2013aln,Aaij:2015esa} decays. The discrepancies observed in these LFD observables with respect to their
%SM predictions are presented in table 1.

The reported observables can be grouped into two sets: $b\to s \mu^+\mu^-$ observables that include only muons,
called as lepton-flavour dependent (LFD) observables, and the other known as lepton-flavour-universality violating
(LFUV) observables that involve both muons and electrons. The set of LFD observables contains several angular
observables, in particular $P_5^{\prime}$ observable in the $B^0\to K^{\ast0}\mu^+\mu^-$ decay \cite{Descotes-Genon:2013vna,Descotes-Genon:2013wba},
showing discrepancies from the SM values, which are collected in table \ref{LFDobs}.
\begin{table*}[!htbp]
\renewcommand{\arraystretch}{1}
	\begin{center}		
		\begin{tabular}{|cccc|}
			\hline
			\makecell{LFD \\ observable} &  Measured value & Deviation & Collaboration   \\
			\hline
              $\langle P_{5}^{\prime}\rangle^{[4.3,\,8.68]}$  &$-0.19^{+0.16}_{-0.16}\pm0.03$        & $3.7\sigma$ & LHCb~\cite{Aaij:2013qta}\\
              $\langle P_{5}^{\prime}\rangle^{[4,\,6]}$       &$-0.300^{+0.158}_{-0.159}\pm0.023$    & $2.8\sigma$ & LHCb~\cite{Aaij:2015oid}               \\
              $\langle P_{5}^{\prime}\rangle^{[6,\,8]}$       &$-0.505^{+0.122}_{-0.122}\pm0.024$    & $3.0\sigma$ &LHCb~\cite{Aaij:2015oid}               \\
              $\langle P_{5}^{\prime}\rangle^{[4,\,6]}$       &$0.26\pm0.35\pm0.18$                  & $2.7\sigma$ &ATLAS~\cite{Aaboud:2018krd}             \\
              $\langle P_{5}^{\prime}\rangle^{[4,\,8]}$      &$-0.267^{+0.275}_{-0.269}\pm0.049$     & $2.1\sigma$ &Belle~\cite{Abdesselam:2016llu}                    \\
			  $\langle P_{5}^{\mu\prime}\rangle^{[4,\,8]}$      &$-0.03^{+0.31}_{-0.30}\pm 0.09$        & $2.6\sigma$ &Belle~\cite{Wehle:2016yoi}                         \\
              $\langle P_{5}^{\prime}\rangle^{[6,\,8.68]}$   &$-0.64^{+0.15}_{-0.19}\pm 0.13$        & $\sim1.0\sigma$ &CMS~\cite{Sirunyan:2017dhj}                      \\
			  $\langle P_{5}^{\prime}\rangle^{[4,\,6]}$      &$-0.439\pm 0.111\pm 0.036$             & $2.5\sigma$ &LHCb~\cite{Aaij:2020nrf}                          \\
			  $\langle P_{5}^{\prime}\rangle^{[6,\,8]}$      & $-0.583\pm 0.090\pm 0.030$            &$2.9\sigma$  &LHCb~\cite{Aaij:2020nrf}                          \\
			\hline					
	    \end{tabular}		
	   \caption{Experimental values for the LFD observable $P_{5}^{\prime}$ in different $q^{2}$ bins.}\label{LFDobs}
	\end{center}
\end{table*}
In addition, more LFD observables such as the branching fractions of the $B\to K\mu^+\mu^-$ \cite{Aaij:2014pli}, $B\to K^{\ast}\mu^+\mu^-$ \cite{Aaij:2014pli,Aaij:2013iag,Aaij:2016flj}, and $B_s\to \phi\mu^+\mu^-$ \cite{Aaij:2013aln,Aaij:2015esa} decays
are found to be on the lower side compared to their SM estimates. These LFD observables, while being sensitive to NP
\cite{Altmannshofer:2008dz,Bobeth:2011nj,Matias:2012xw,DescotesGenon:2012zf,Matias:2014jua}, can not establish the NP
case unambiguously because of the involvement of the hadronic uncertainties originating from the different long-distance
effects, in particular from form factors, power corrections, and charm resonances \cite{Khodjamirian:2010vf,Khodjamirian:2012rm,
Lyon:2014hpa,Descotes-Genon:2014uoa,Capdevila:2017ert,Blake:2017fyh}. Therefore, without having
additional data or a complete and reliable calculations of the hadronic uncertainties there remains a possibility to explain
the currently observed LFD anomalies with more conservative assumptions on the involved hadronic contributions
\cite{Jager:2012uw,Jager:2014rwa,Ciuchini:2015qxb,Ciuchini:2016weo,Bobeth:2017vxj}.
%remains a possibility that these contributions may well be the ultimate cause behind the currently observed LFD anomalies.
%Therefore, without having a complete and reliable calculations of these contributions there remains a possibility that
%within the SM the observed LFD anomalies could be accounted for within a SM picture with underestimated hadronic uncertainties.

The second set with LFUV observables includes the ratios of branching fractions involving both $b\to s \mu^+\mu^-$ and
$b\to s e^+e^-$ transitions. In table \ref{LFUVratios}, we list the deviations observed by the LHCb collaboration
in ratios $R_K\equiv \mathcal{B}(B^+\to K^+\mu^+\mu^-)/\mathcal{B}(B^+\to K^+e^+e^-)$, and $R_{K^{\ast}}\equiv \mathcal{B}(B^0\to K^{\ast0}\mu^+\mu^-)/\mathcal{B}(B^0\to K^{\ast0}e^+e^-)$
from their SM expectation of $\simeq1$ \cite{Hiller:2003js,Bordone:2016gaq}. Additionally, we have recent Belle results for
$R_K$ \cite{BELLE:2019xld} and $R_{K^{\ast}}$ \cite{Belle:2019oag}, which are combined together for the charged and
neutral decay modes, and are presented in multiple $q^2$ bins.
\begin{table*}[!htbp]
\renewcommand{\arraystretch}{1}
	\begin{center}
		  \begin{tabular}{|lllc|}
			\hline
			\makecell{LFUV \\ observable} &  \multicolumn{2}{c}{Measured value}   & Deviation \\
			\hline
				$R_{K}^{\,[1,\,6]}$              & $0.745_{-0.074}^{+0.090}\pm0.036$        &\cite{Aaij:2014ora}        & $2.6\sigma$\\
				$R_{K}^{\,[1.1,\,6]}$            & $0.846^{+0.060+0.016}_{-0.054-0.014}$    &\cite{Aaij:2019wad}        & $2.5\sigma$\\
				$R_{K}^{\,[1.1,\,6]}$            & $0.846^{+0.042+0.013}_{-0.039-0.012}$    &\cite{LHCb:2021trn}        & $3.1\sigma$\\
				$R_{K^{\ast}}^{\,[0.045,\,1.1]}$ & $0.66_{-0.07}^{+0.11}\pm0.03$            &\cite{Aaij:2017vbb}        & $2.4\sigma$\\
				$R_{K^{\ast}}^{\,[1.1,\,6]}$     & $0.69_{-0.07}^{+0.11}\pm0.05$            &\cite{Aaij:2017vbb}        & $2.5\sigma$\\
			\hline
		  \end{tabular}
        \caption{LHCb predictions for the LFUV ratios in different $q^{2}$ bins.}		
		\label{LFUVratios}
	\end{center}
\end{table*}
However, due to large errors, these results are in agreement with both the SM and the LHCb measurements. Moreover, additional LFUV observables, such as
$Q_{4,5}=P_{4,5}^{\mu\prime}-P_{4,5}^{e\prime}$ \cite{Capdevila:2016ivx}, have been observed by the Belle collaboration
\cite{Wehle:2016yoi}. Furthermore, LHCb has also performed the test of lepton flavour universality (LFU) violation in the baryon decay $\Lambda_b \to p K^-\ell^+\ell^-$ \cite{Aaij:2019bzx}, and the decays $B^+\to {K^*}^+\ell^+\ell^-$ and $B^0\to {K_S}^0\ell^+\ell^-$~\cite{LHCb:2021lvy} which are isospin partners of the formerly tested $B^0\to {K^*}^0\ell^+\ell^-$ and $B^+\to {K}^+\ell^+\ell^-$ decays. All the measured central values of the LFUV ratios corresponding to these decays are lower than the SM predictions, which shows a consistent tendency. Contrary to the LFD observables, SM predictions for the LFUV observables $R_K$ and $R_{K^{\ast}}$ are theoretically clean as the hadronic uncertainties essentially cancel and therefore they hold the key to unravel NP without ambiguity.

%Because all (LFD+LFUV) observables involve $b\to s \mu^+\mu^-$ transition and additionally LFUV NP contributions are mandatory
%to account for the LFUV anomalies, it seems natural to address all the $b\to s \ell^+\ell^-$ data by assuming NP present only in
%the $b\to s \mu^+\mu^-$ sector, which automatically breaks lepton flavour universality and is thus purely LFUV.
Interestingly, several model independent global fit analyses \cite{Alguero:2021anc,Descotes-Genon:2015uva,Altmannshofer:2017fio,Alok:2017sui,Altmannshofer:2017yso,
Geng:2017svp,Ciuchini:2017mik,Capdevila:2017bsm,Alguero:2019ptt,Alok:2019ufo,Ciuchini:2019usw,Datta:2019zca,Aebischer:2019mlg,
Kowalska:2019ley,Arbey:2019duh,Bhattacharya:2019dot,Biswas:2020uaq} performed with the assumption
of LFUV NP present only in the $b\to s \mu^+\mu^-$ sector have pointed out two simple
one-dimensional (1D) NP scenarios (S1) $C_{9\mu}^{\text{NP}}$ or (S2) $C_{9\mu}^{\text{NP}}=-C_{10\mu}^{\text{NP}}$, that can provide
better fit to all the $b\to s$ data with preferences reaching $\approx5-6\sigma$ compared to the SM. However, performing the separate fits,
it is observed that the inclusion of the more recent $R_K$ and $R_{K^{\ast}}$ data has created tensions between the separate fit to LFD and
LFUV set in both the scenarios S1 and S2 along with increasing the significance gap between the two LFUV fits of the two scenarios
\cite{Alguero:2021anc,Alguero:2019ptt,Datta:2019zca}. These tensions, if not statistical fluctuations, could be indications of NP also present
in $b\to s e^+e^-$. For instance, in Ref. \cite{Datta:2019zca}, it is shown that the additional LFUV NP in $b\to s e^+e^-$
along with the basic scenarios S1 and S2 leads to a number of new scenarios, which can remove tensions along with improving the
overall fit to all data. Another complementary approach proposed in \cite{Alguero:2018nvb}, before the latest $R_K$ and
$R_{K^{\ast}}$ measurement, showed that several scenarios with both LFU and LFUV NP contributions to the Wilson
coefficients (WC), $C_{9\ell}^{\text{NP}}$ and $C_{10\ell}^{\text{NP}}$ can improve the agreement with the overall data.
After the latest $R_K$ and $R_{K^{\ast}}$ data, more updated global fit analyses \cite{Alguero:2021anc,Alguero:2019ptt}, have again
pointed out various NP scenarios with enhanced significance, and an improved preference
for the NP scenarios with right-handed currents (RHC) have been observed to emerge. Furthermore, a better
description of data can also be obtained by increasing the degrees of freedom, i.e., 2D fits, along with the assumptions
such as NP affects only muons.

As there is no unique solution and many new scenarios are piling up due to the recently emerging NP patterns in global fit
analyses, it is particularly important to discriminate between different possible scenarios and to devise methods to further
confirm or constrain patterns of NP \cite{Kumbhakar:2018uty,Alok:2020bia}. One way is to consider the complementary channels
induced by the same quark level $b\to s \ell^+\ell^-$ transitions, and analyze the implications of the different NP scenarios
for the theoretically clean LFUV ratios in different kinematical regions. The list of decay channels
induced by the $b\to s \ell^+\ell^-$ transition is long, and the LFUV ratios in a number of
decay channels have been studied \cite{Hiller:2014ula,Wang:2017mrd,Huang:2018rys,Dutta:2019wxo} based on the previous data. In the present study, we consider the most recent experimental results and restrict to seven exclusive channels
$M_{in}\to M_{f} \ell^+\ell^-$, with $M_{in}=B, B_s, \Lambda_b$ and
$M_{f}=f_0, K_0^{\ast}, K, K^{\ast}, \phi, K_1, \Lambda$. We study these channels in the model independent effective Hamiltonian
approach by employing the helicity formalism. The theoretical analysis of LFUV observables in complementary
hadronic decays, along with providing additional tests of LFU, similar to being performed in $R_K$ and $R_{K^{\ast}}$,
can help to distinguish and further strengthen some of the emerging NP patterns in the global fit analyses, before going on to
build the accurate NP models accommodating the $B$ decay anomalies \cite{Alonso:2017bff,Alonso:2017uky,Duan:2018akc}.
%The theoretical analysis of LFUV observables in complementary
%hadronic decays can help to distinguish and further strengthen the emerging NP patterns along with
%providing interesting cross checks.

The paper is organized as follows. In section \ref{effH}, we describe the effective Hamiltonian and the
decay amplitude for $b\to s \ell^+\ell^-$ transitions. In section \ref{secHelicity}, we consider the helicity
formalism and work out the explicit expressions of the helicity amplitudes
for the considered decays. In section \ref{physobs}, we construct LFUV observables. Section \ref{secNum}, is devoted to numerical analysis, where
we also present our choice of the NP scenarios from different global fit analyses. The results are
summarized in section \ref{summary}.
%Details of the SM Wilson coefficients, hadronic matrix elements, numerical inputs, and the technical
%considerations concerning the kinematics and polarization conventions, along with the predictions of the LFUV ratios
%in the SM and the NP scenarios are collected in appendices.
\section{Effective Hamiltonian and decay amplitude}\label{effH}
The effective weak Hamiltonian for $b\to s\ell^{+}\ell^{-}$ transition is given by
\begin{align}\label{H1}
\mathcal{H}_{\text{eff}}=-\frac{4 G_{F}}{\sqrt{2}}V_{tb}V^{\ast}_{ts}&\Bigg[\sum_{i=1}^{6}C_{i}O_{i}
+\sum_{i=7}^{8}\big(C_{i}O_{i}+C_{i^{\prime}}O_{i^{\prime}}\big)\notag\\
&+\sum_{i=9,10}\Big((C_{i}+C_{i\ell}^{\text{NP}})O_{i}+C_{i^{\prime}\ell}^{\text{NP}}O_{i^{\prime}}\Big)\Bigg],
\end{align}
where we have neglected the doubly Cabibbo suppressed contribution $(\propto V_{ub}V_{us}^\ast)$, and
$G_F$ is the Fermi coupling constant. The operators $O_{i\le6}$ are the same as the $P_{1,2}^c, P_{3,...,6}^{}$,
given in Ref. \cite{Bobeth:1999mk}, and the others are
\begin{align}\label{op1}
O_{7} &=\frac{e}{16\pi ^{2}}m_{b}\left( \bar{s}\sigma _{\mu \nu }P_{R}b\right) F^{\mu \nu },
&  O_{7^{\prime}} &=\frac{e}{16\pi ^{2}}m_{b}\left( \bar{s}\sigma _{\mu \nu }P_{L}b\right) F^{\mu \nu },\notag\\
O_{8} &=\frac{g_s}{16\pi ^{2}}m_{b}\left( \bar{s}\sigma _{\mu \nu } T^a P_{R}b\right) G^{\mu \nu\,a},
&  O_{8^{\prime}} &=\frac{g_s}{16\pi ^{2}}m_{b}\left( \bar{s}\sigma _{\mu \nu } T^a P_{L}b\right) G^{\mu \nu\,a},\notag\\
O_{9} &=\frac{e^{2}}{16\pi ^{2}}(\bar{s}\gamma _{\mu }P_{L}b)(\bar{l}\gamma^{\mu }l),
&  O_{9^{\prime}} &=\frac{e^{2}}{16\pi ^{2}}(\bar{s}\gamma _{\mu }P_{R}b)(\bar{l}\gamma^{\mu }l),\notag\\
O_{10} &=\frac{e^{2}}{16\pi ^{2}}(\bar{s}\gamma _{\mu }P_{L}b)(\bar{l} \gamma ^{\mu }\gamma _{5} l),
&  O_{10^{\prime}} &=\frac{e^{2}}{16\pi ^{2}}(\bar{s}\gamma _{\mu }P_{R}b)(\bar{l} \gamma ^{\mu }\gamma _{5} l),
\end{align}
where $e$ $(g_s)$ is the electromagnetic (strong) coupling constant, and $m_b$ represents the running
$b-$quark mass in the $\overline{\text{MS}}$ scheme.
%In Eq.(\ref{H1}) $C_{i}$'s are the scale dependent Wilson coefficients which incorporates the effects of high energy physics and
%can be evaluated via perturbative approach. However the local quark and electromagnetic operators $O_{i}$'s contain the non-perturbative
%long distance physics.

Within the SM, major role in $b\to s\ell^{+}\ell^{-}$ transition, is played by operators $O_{7,9,10}$, whereas contributions of
primed dipole operators $O_{7^{\prime},8^{\prime}}$ are suppressed by $m_s/m_b$, and therefore we neglect them. Furthermore,
the factorizable contributions from current-current, QCD penguins and chromomagnetic dipole operators $O_{1-6,8}$ can be absorbed
into the effective Wilson coefficients $C_{7}^{\text{eff}}(q^{2})$ and $C_{9}^{\text{eff}}(q^{2})$ \cite{Bobeth:1999mk,Beneke:2001at,Asatrian:2001de,Asatryan:2001zw,Greub:2008cy,Du:2015tda}.
The explicit expressions of these Wilson coefficients, which we used, are presented in appendix \ref{append}. It is important to mention that, in
Eq. (\ref{H1}), we have considered NP contributions only in $O_{9^{(\prime)}}$ and $O_{10^{(\prime)}}$ operators because the emerging
viable NP solutions from the global fits of all the $b\to s$ data, which we consider in our study, are only
in the form of vector and axial-vector operators. The numerical values of Wilson coefficients at $\mu\sim m_{b}$
in the SM are presented in table \ref{wc table}.
%\begin{table*}[ht]
%\centering
%\begin{adjustbox}{max width=\textwidth}
%\begin{tabular}{cccccccccc}
%\hline\hline
%$C_{1}$&$C_{2}$&$C_{3}$&$C_{4}$&$C_{5}$&$C_{6}$&$C_{7}$&$C_{9}$&$C_{10}$
%\\ \hline
%  \ \ -0.263 \ \ &  \ \ 1.011 \ \ & \ \ 0.005 \ \ &  \ \ -0.0806  \ \ &
%  \ \ 0.0004 \ \ &  \ \ 0.0009  \ \ &  \ \ -0.2923 \ \ &  \ \ 4.0749 \ \ & \ \ -4.3085\ \ \ \\
%\hline\hline
%\end{tabular}
%\end{adjustbox}
%\caption{The Wilson coefficients $C_{i}^{\mu}$ at the scale $\mu\sim m_{b}$ in the SM.}
%\label{wc table}
%\end{table*}
%\begin{table*}[t]
%\setlength\tabcolsep{0pt}
%\centering
%\begin{adjustbox}{max width=0.95\textwidth}
%\begin{tabular}{|cccccccccc|}
%\hline
%$C_{1}$&$C_{2}$&$C_{3}$&$C_{4}$&$C_{5}$&$C_{6}$&$C_{7}$&$C_{8}$&$C_{9}$&$C_{10}$
%\\ \hline
%  \ \ $-0.294$ \ \ &  \ \ $1.017$ \ \ & \ \ $-0.0059$ \ \ &  \ \ $-0.087$  \ \ &
%  \ \ $0.0004$ \ \ &  \ \ $0.0011$  \ \ &  \ \ $-0.324$ \ \ & \ \ $-0.176$ \ \ &
%   \ \ $4.114$ \ \ & \ \ $-4.193$\ \ \ \\
%\hline
%\end{tabular}
%\end{adjustbox}
%\caption{The SM Wilson coefficients $C_{i}^{\mu}$ up to NNLL accuracy given at the scale $\mu\sim m_{b}$.}
%\label{wc table}
%\end{table*}
\begin{table*}[t]
\centering
\begin{tabular}{|cccccccccc|}
\hline
$C_{1}$&$C_{2}$&$C_{3}$&$C_{4}$&$C_{5}$&$C_{6}$&$C_{7}$&$C_{8}$&$C_{9}$&$C_{10}$
\\ \hline
  $-0.294$ &   $1.017$  & $-0.0059$  &   $-0.087$  &
  $0.0004$  &  $0.0011$   &   $-0.324$  &  $-0.176$  &
    $4.114$  &  $-4.193$ \\
\hline
\end{tabular}
\caption{The SM Wilson coefficients $C_{i}^{\mu}$ up to NNLL accuracy given at the scale $\mu\sim m_{b}$.}
\label{wc table}
\end{table*}

Using the effective Hamiltonian given in Eq. (\ref{H1}), the decay amplitude for the process $M_{in}\to M_{f}\ell^{+}\ell^{-}$,
including the SM and the NP contributions, can be written as\footnote{We neglect the non-factorizable contributions such as the non-perturbative charm-loop corrections which are not the expected sources of the deviations in $R_{K^{(*)}}$~\cite{Capdevila:2017ert,Khodjamirian:2010vf}.}
\begin{eqnarray}
\mathcal{M}\left(M_{in}\to M_{f}\ell^{+}\ell^{-}\right)=\frac{G_{F}\alpha}{2\sqrt{2}\pi}V_{tb}V^{\ast}_{ts}\Big\{T^{1,M_f}_{\mu}(\bar{\ell}\gamma^{\mu}\ell)
+T^{2,M_f}_{\mu}(\bar{\ell}\gamma^{\mu}\gamma_{5}\ell)\Big\},\label{Amp1}
\end{eqnarray}
where
\begin{eqnarray}
T^{1,M_f}_{\mu}&=&(C_{9}^{\text{eff}}+C_{9\ell}^{\text{NP}})\Big\langle M_{f}(k)|\bar s\gamma_{\mu}(1-\gamma_{5})b|M_{in}(p)\Big\rangle
+C_{9^\prime\ell}^{\text{NP}}\Big\langle M_{f}(k)|\bar s\gamma_{\mu}(1+\gamma_{5})b|M_{in}(p)\Big\rangle\notag\\
&-&\frac{2m_{b}}{q^{2}}C_{7}^{\text{eff}}
\Big\langle M_{f}(k)|\bar s i\sigma_{\mu\nu}q^{\nu}(1+\gamma_{5})b|M_{in}(p)\Big\rangle,\label{Amp1a}
\\
T^{2,M_f}_{\mu}&=&(C_{10}+C_{10\ell}^{\text{NP}})\Big\langle M_{f}(k)|\bar s\gamma_{\mu}(1-\gamma_{5})b|M_{in}(p)\Big\rangle
+C_{10^\prime\ell}^{\text{NP}}\Big\langle M_{f}(k)|\bar s\gamma_{\mu}(1+\gamma_{5})b|M_{in}(p)\Big\rangle.\label{Amp1b}
\end{eqnarray}
%\begin{eqnarray}
%T^{1,M_f}_{\mu}&=&\Big\{(C_{9}^{\text{eff}}+C_{9\ell}^{\text{NP}})\langle M_{f}(k)|\bar s\gamma_{\mu}(1-\gamma_{5})b|M_{in}(p)\rangle\notag
%\\
%&+&C_{9^\prime\ell}^{\text{NP}}\langle M_{f}(k)|\bar s\gamma_{\mu}(1+\gamma_{5})b|M_{in}(p)\rangle\notag\\
%&-&\frac{2m_{b}}{q^{2}}C_{7}^{\text{eff}}
%\langle M_{f}(k)|\bar s i\sigma_{\mu\nu}q^{\nu}(1+\gamma_{5})b|M_{in}(p)\rangle\Big\},\label{Amp1a}
%\\
%T^{2,M_f}_{\mu}&=&\Big\{(C_{10}+C_{10\ell}^{\text{NP}})\langle M_{f}(k)|\bar s\gamma_{\mu}(1-\gamma_{5})b|M_{in}(p)\rangle\notag\\
%&+&C_{10^\prime\ell}^{\text{NP}}\langle M_{f}(k)|\bar s\gamma_{\mu}(1+\gamma_{5})b|M_{in}(p)\rangle\Big\}.\label{Amp1b}
%\end{eqnarray}
%%\section{Hadronic matrix elements and decay amplitudes}\label{matelem}
To calculate $T^{i,M_f}_{\mu}$ $(i=1, 2)$, one requires the involved hadronic matrix elements which can be parameterized in terms of the transition form factors. As we consider various decay channels with $M_{in}=B, B_s, \Lambda_b$ and $M_{f}=f_0, K_0^{\ast}, K, K^{\ast}, \phi, K_1, \Lambda$, we give the hadronic matrix elements, in terms of the transition form factors, for each case in appendix~\ref{app:HME}. The form factors, for the decay $B_{s}\to f_{0}(980)\ell^{+}\ell^{-}$, and $B\to K^{\ast}_0(1430)\ell^{+}\ell^{-}$ can be calculated using the light cone QCD sum rule approach \cite{Colangelo:2010bg}, and three-point QCD sum rules \cite{Aliev:2007rq}. For $B\to K$ transition form factors, light cone sum rules (LCSR)
predictions can be extrapolated at $q^{2}\leq 8\text{GeV}^{2}$ to the whole
kinematical region by applying  $z$-series expansion \cite{Lu:2018cfc}. The simplified series
expansion for $B\to P$ form factors has been adopted which was originally proposed in \cite{Bourrely:2008za}.
For the decays $B\to K^{\ast}\ell^{+}\ell^{-}$ and $B_{s}\to\phi\ell^{+}\ell^{-}$,
we use the series expansion fits to LCSR and lattice form factors \cite{Straub:2015ica}. The transition
form factors in terms of rapidly converging series parameter can be expressed as \cite{Straub:2015ica}
$F_{i}(q^{2})=P_{i}(q^{2})\sum_{k}\alpha^{i}_{k}[z^{\prime}(q^{2})-z^{\prime}(0)]^{k}$,
where $P_{i}(q^{2})=\frac{1}{(1-q^{2}/m^{2}_{R,i})}$ is simple pole representing the first resonance in the spectrum.
For $B\to K_1(1270,1400)\ell^+\ell^-$ decay, the physical states $K_{1}(1270)$ and $K_{1}(1400)$ are mixed
states of $K_{1A}$ and $K_{1B}$ with mixing angle $\theta_{K_1}$ defined as
\begin{eqnarray}
|K_{1}(1270)\rangle&=&|K_{1A}\rangle\sin\theta_{K_{1}}+|K_{1B}\rangle\cos\theta_{K_{1}}\,,\label{theta}\\
|K_{1}(1400)\rangle&=&|K_{1A}\rangle\cos\theta_{K_{1}}-|K_{1B}\rangle\sin\theta_{K_{1}}\,.\label{theta1}
\end{eqnarray}
The corresponding mixing relations among different matrix elements and for the form factors are explicitly
given in \cite{Hatanaka:2008gu,Ishaq:2013toa,Huang:2018rys,Munir:2015gsp}. For numerical analysis, we use the light-cone QCD sum rule form factors \cite{Hatanaka:2008gu}. For the $\Lambda_b\to \Lambda\ell^{+}\ell^{-}$ decay, we use the lattice QCD results of the form factors for whole $q^2$ range \cite{Detmold:2016pkz}. The form factors used in \cite{Detmold:2016pkz} are related to our notation of the form factors as $f^V_{t, 0, \perp}=f_{0, +, \perp}$,
$f^A_{t, 0, \perp}=g_{0, +, \perp}$, $f^T_{0, \perp}=h_{+, \perp}$, and $f^{T_5}_{0, \perp}=\tilde h_{+, \perp}$.
%In the numerical analysis, to estimate the central value of the LFUV ratio and the total uncertainty due to the form factors, we use the ``nominal'' fit and ``higher-order'' fit form factors and employ the steps given in Eqs. (50)-(55), in
%Ref. \cite{Detmold:2016pkz}.
%\subsection{Helicity amplitudes}
\section{Helicity formalism and helicity amplitudes}\label{secHelicity}
The decay amplitudes can be expressed in terms of helicity basis as described in \cite{Faessler:2002ut},
and references therein. The orthonormality and completeness properties of helicity basis $\varepsilon^{\alpha}(n=t, +, -, 0)$, with three
spin $1$ components orthogonal to momentum transfer i.e., $q\cdot\varepsilon(\pm)=q\cdot\varepsilon=0$, can be expressed as follows
\begin{eqnarray}
\varepsilon^{\ast\alpha}(n)\varepsilon_{\alpha}(l)=g_{nl}, \qquad\quad \sum_{n, l=t, +, -, 0}\varepsilon^{\ast\alpha}(n)\varepsilon^{\beta}(l)g_{nl}=g^{\alpha\beta},\label{C22}
\end{eqnarray}
with $g_{nl}=\text{diag}(+, -, -, -)$. Using the completeness property given in Eq. (\ref{C22}), the contraction of leptonic tensors $L^{(k)\alpha\beta}$ and hadronic tensors $H^{ij}_{\alpha\beta}=T^{i,M_f}_{\alpha}\overline{T}^{\,j,M_f}_{\beta}$ $(i, j=1, 2)$, can be written as
\begin{eqnarray}
L^{(k)\alpha\beta}H^{ij}_{\alpha\beta}=\sum_{n, n^{\prime}, l, l^{\prime}}L^{(k)}_{nl}g_{nn^{\prime}}g_{ll^{\prime}}H^{ij}_{n^{\prime}l^{\prime}},\label{LH}
\end{eqnarray}
where the leptonic and hadronic tensors are expressed in the helicity basis as follows
\begin{eqnarray}
L^{(k)}_{nl}=\varepsilon^{\alpha}(n)\varepsilon^{\ast\beta}(l)L^{(k)}_{\alpha\beta}, &&\qquad H^{ij}_{nl}=\varepsilon^{\ast\alpha}(n)\varepsilon^{\beta}(l)H^{ij}_{\alpha\beta}.\label{LHT}
\end{eqnarray}
Both leptonic and hadronic tensors given in Eq. (\ref{LHT}), will be evaluated in two different frame of references. The lepton
tensor $L^{(k)}_{nl}$ will be evaluated in $l\bar{l}$ CM frame. However the hadron tensor $H^{ij}_{nl}$ will be evaluated
in the rest frame of decaying hadron.
\subsection{Helicity amplitudes for $M_{in}\to S\ell^{+}\ell^{-}$ decays}
\begin{eqnarray}
H^{ij}_{nl}=\big(\varepsilon^{\ast\alpha}(n)T^{i,S}_{\alpha}\big)\cdot\big(\overline{\varepsilon^{\ast\beta}(l)T^{j,S}_{\beta}}\big)
\equiv H^{i,S}_n \,\overline{H}^{\,j,S}_l,\label{HA1}
\end{eqnarray}
where, for $M_{in}=B_s$, and $S=f_0(980)$, explicit helicity amplitudes are obtained as
\begin{eqnarray}
H^{1,f_0}_t&=&i\frac{m^2_{B_s}-m^2_{f_0}}{\sqrt{q^2}}(C_{9}^{\text{eff}}+C_{9\ell}^{\text{NP}}-C_{9^{\prime}\ell}^{\text{NP}})f^{f_0}_0(q^2),\notag
\\
H^{2,f_0}_t&=&i\frac{m^2_{B_s}-m^2_{f_0}}{\sqrt{q^2}}(C_{10}+C_{10\ell}^{\text{NP}}-C_{10^{\prime}\ell}^{\text{NP}})f^{f_0}_0(q^2),\notag
\\
H^{i,f_0}_{\pm}&=&0,\notag
\\
H^{1,f_0}_0&=&i\sqrt{\frac{\lambda}{q^2}}\Big[(C_{9}^{\text{eff}}+C_{9\ell}^{\text{NP}}-C_{9^{\prime}\ell}^{\text{NP}})f^{f_0}_+(q^2)
+\frac{2m_b}{m_{B_s}+m_{f_0}}C_{7}^{\text{eff}}f^{f_0}_T(q^2)\Big],\notag
\\
H^{2,f_0}_0&=&i\sqrt{\frac{\lambda}{q^2}}\Big[(C_{10}+C_{10\ell}^{\text{NP}}-C_{10^{\prime}\ell}^{\text{NP}})f^{f_0}_+(q^2)\Big],\label{HA2}
\end{eqnarray}
similarly, for $M_{in}=B$, and $S=K^{\ast}_0(1430)$,
\begin{eqnarray}
H^{1,K^{\ast}_0}_t&=&i(C_{9}^{\text{eff}}+C_{9\ell}^{\text{NP}}-C_{9^{\prime}\ell}^{\text{NP}})
\Big[\frac{m^2_{B}-m^2_{K^{\ast}_0}}{\sqrt{q^2}}f^{K^{\ast}_0}_+(q^2)+\sqrt{q^2}f^{K^{\ast}_0}_-(q^2)\Big],\notag
\\
H^{2,K^{\ast}_0}_t&=&i(C_{10}+C_{10\ell}^{\text{NP}}-C_{10^{\prime}\ell}^{\text{NP}})
\Big[\frac{m^2_{B}-m^2_{K^{\ast}_0}}{\sqrt{q^2}}f^{K^{\ast}_0}_+(q^2)+\sqrt{q^2}f^{K^{\ast}_0}_-(q^2)\Big],\notag
\\
H^{i,K^{\ast}_0}_{\pm}&=&0,\notag
\\
H^{1,K^{\ast}_0}_0&=&i\sqrt{\frac{\lambda}{q^2}}\Big[(C_{9}^{\text{eff}}+C_{9\ell}^{\text{NP}}-C_{9^{\prime}\ell}^{\text{NP}})f^{K^{\ast}_0}_+(q^2)
+\frac{2m_b}{m_{B}+m_{K^{\ast}_0}}C_{7}^{\text{eff}}f^{K^{\ast}_0}_T(q^2)\Big],\notag
\\
H^{2,K^{\ast}_0}_0&=&i\sqrt{\frac{\lambda}{q^2}}\Big[(C_{10}+C_{10\ell}^{\text{NP}}-C_{10^{\prime}\ell}^{\text{NP}})f^{K^{\ast}_0}_+(q^2)\Big].\label{HA3}
\end{eqnarray}
Here $\lambda\equiv \lambda(m^2_{B_s(B)}, m^2_{f_0(K^{\ast}_0)}, q^2)$.
\subsection{Helicity amplitudes for $M_{in}\to P\ell^{+}\ell^{-}$ decays}
\begin{eqnarray}
H^{ij}_{nl}=\big(\varepsilon^{\ast\alpha}(n)T^{i,P}_{\alpha}\big)\cdot\big(\overline{\varepsilon^{\ast\beta}(l)T^{j,P}_{\beta}}\big)
\equiv H^{i,P}_n \,\overline{H}^{\,j,P}_l,\label{HAA11}
\end{eqnarray}
where, for $M_{in}=B$, and $P=K$, explicit helicity amplitudes are calculated as
\begin{eqnarray}
H^{1,K}_t&=&\frac{m^2_{B}-m^2_{K}}{\sqrt{q^2}}(C_{9}^{\text{eff}}+C_{9\ell}^{\text{NP}}+C_{9^{\prime}\ell}^{\text{NP}})f^{K}_0(q^2),\notag
\\
H^{2,K}_t&=&\frac{m^2_{B}-m^2_{K}}{\sqrt{q^2}}(C_{10}+C_{10\ell}^{\text{NP}}+C_{10^{\prime}\ell}^{\text{NP}})f^{K}_0(q^2),\notag
\\
H^{i,K}_{\pm}&=&0,\notag
\\
H^{1,K}_0&=&\sqrt{\frac{\lambda}{q^2}}\Big[(C_{9}^{\text{eff}}+C_{9\ell}^{\text{NP}}+C_{9^{\prime}\ell}^{\text{NP}})f^{K}_+(q^2)
+\frac{2m_b}{m_{B}+m_{K}}C_{7}^{\text{eff}}f^{K}_T(q^2)\Big],\notag
\\
H^{2,K}_0&=&\sqrt{\frac{\lambda}{q^2}}\Big[(C_{10}+C_{10\ell}^{\text{NP}}+C_{10^{\prime}\ell}^{\text{NP}})f^{K}_+(q^2)\Big].\label{HA4}
\end{eqnarray}
Here $\lambda\equiv \lambda(m^2_{B}, m^2_{K}, q^2)$.
\subsection{Helicity amplitudes for $M_{in}\to V\ell^{+}\ell^{-}$ decays}
\begin{eqnarray}
H^{ij}_{nl}&=&\big(\varepsilon^{\ast\alpha}(n)T^{i,V}_{\alpha}\big)\cdot\big(\overline{\varepsilon^{\ast\beta}(l)T^{j,V}_{\beta}}\big)\notag
\\
&=&\big(\varepsilon^{\ast\alpha}(n)\overline\epsilon^{\ast\mu}(r)T^{i,V}_{\alpha,\mu}\big)\cdot\big(\overline{\varepsilon^{\ast\beta}(l)
\overline\epsilon^{\ast\nu}(s)T^{j,V}_{\beta,\nu}}\big)\delta_{rs}\equiv H^{i,V}_n \,\overline{H}^{\, j,V}_l,\label{HA5}
\end{eqnarray}
where, from angular momentum conservation, $r=n$ and $s=l$ for $n, l=\pm,0$ and $r, s=0$ for $n, l=t$.
The explicit helicity amplitudes for $M_{in}=B(B_s)$, and $V=K^{\ast}(\phi)$, are derived in terms of the Wilson coefficients and the form factors
as~\footnote{With the different conventions used in \cite{Gratrex:2015hna,Ebert:2010dv}, similar expressions of the helicity amplitudes are obtained for $B\to K^*$ channel by employing the more sophisticated generalized helicity amplitude formalism.}
\begin{align}
H^{1,K^{\ast}(\phi)}_t&=-i\sqrt{\frac{\lambda}{q^2}}(C_{9}^{\text{eff}}+C_{9\ell}^{\text{NP}}-C_{9^{\prime}\ell}^{\text{NP}})A^{K^{\ast}(\phi)}_0,\notag
\\
H^{2,K^{\ast}(\phi)}_t&=-i\sqrt{\frac{\lambda}{q^2}}(C_{10}+C_{10\ell}^{\text{NP}}-C_{10^{\prime}\ell}^{\text{NP}})A^{K^{\ast}(\phi)}_0,\notag
\\
H^{1,K^\ast(\phi)}_{\pm}&=-i\left(m^2_{B(B_s)}-m^2_{K^\ast(\phi)}\right)\Big[(C_{9}^{\text{eff}}+C_{9\ell}^{\text{NP}}-C_{9^{\prime}\ell}^{\text{NP}})
\frac{A_{1}^{K^\ast(\phi)}}{\left(m_{B(B_s)}-m_{K^\ast(\phi)}\right)}\notag
\\
&+\frac{2m_{b}}{q^{2}}C_{7}^{\text{eff}}T_{2}^{K^\ast(\phi)}\Big]
\pm i\sqrt{\lambda}\Big[(C_{9}^{\text{eff}}+C_{9\ell}^{\text{NP}}+C_{9^{\prime}\ell}^{\text{NP}})
\frac{V^{K^\ast(\phi)}}{\left(m_{B(B_s)}+m_{K^\ast(\phi)}\right)}\notag
\\
&\qquad\quad+\frac{2m_{b}}{q^{2}}C_{7}^{\text{eff}}T_{1}^{K^\ast(\phi)}\Big],\notag
\\
H^{2,K^\ast(\phi)}_{\pm}&=-i(C_{10}+C_{10\ell}^{\text{NP}}-C_{10^{\prime}\ell}^{\text{NP}})\left(m_{B(B_s)}+m_{K^\ast(\phi)}\right)
A_{1}^{K^\ast(\phi)}\notag
\\
&\pm i\sqrt{\lambda}(C_{10}+C_{10\ell}^{\text{NP}}+C_{10^{\prime}\ell}^{\text{NP}})
\frac{V^{K^\ast(\phi)}}{\left(m_{B(B_s)}+m_{K^\ast(\phi)}\right)},\notag
\\
H^{1,K^\ast(\phi)}_0&=-\frac{8im_{B(B_s)}m_{K^\ast(\phi)}}{\sqrt{q^2}}\Bigg[(C_{9}^{\text{eff}}+C_{9\ell}^{\text{NP}}-C_{9^{\prime}\ell}^{\text{NP}})
A_{12}^{K^\ast(\phi)}+m_b C_{7}^{\text{eff}}\frac{T_{23}^{K^\ast(\phi)}}{m_{B(B_s)}+m_{K^\ast(\phi)}}\Bigg],\notag
\\
H^{2,K^\ast(\phi)}_0&=-\frac{8im_{B(B_s)}m_{K^\ast(\phi)}}{\sqrt{q^2}}\Bigg[(C_{10}+C_{10\ell}^{\text{NP}}-C_{10^{\prime}\ell}^{\text{NP}})
A_{12}^{K^\ast(\phi)}\Bigg].\label{HA6}
\end{align}
Here $\lambda\equiv \lambda(m^2_{B(B_s)}, m^2_{K^\ast(\phi)}, q^2)$.
\subsection{Helicity amplitudes for $M_{in}\to A\ell^{+}\ell^{-}$ decays}
\begin{eqnarray}
H^{ij}_{nl}&=&\big(\varepsilon^{\ast\alpha}(n)T^{i,A}_{\alpha}\big)\cdot\big(\overline{\varepsilon^{\ast\beta}(l)T^{j,A}_{\beta}}\big)\notag
\\
&=&\big(\varepsilon^{\ast\alpha}(n)\overline\epsilon^{\ast\mu}(r)T^{i,A}_{\alpha,\mu}\big)\cdot\big(\overline{\varepsilon^{\ast\beta}(l)
\overline\epsilon^{\ast\nu}(s)T^{j,A}_{\beta,\nu}}\big)\delta_{rs}\equiv H^{i,A}_n \,\overline{H}^{\, j,A}_l,\label{HA7}
\end{eqnarray}
where, from angular momentum conservation, $r=n$ and $s=l$ for $n, l=\pm,0$ and $r, s=0$ for $n, l=t$.
The explicit helicity amplitudes for $M_{in}=B$, and $A=K_1$, are given as
\begin{align}
H^{1,K_1}_t&=-\sqrt{\frac{\lambda}{q^2}}(C_{9}^{\text{eff}}+C_{9\ell}^{\text{NP}}+C_{9^{\prime}\ell}^{\text{NP}})V^{K_1}_0,\notag
\\
H^{2,K_1}_t&=-\sqrt{\frac{\lambda}{q^2}}(C_{10}+C_{10\ell}^{\text{NP}}+C_{10^{\prime}\ell}^{\text{NP}})V^{K_1}_0,\notag
\end{align}
\begin{align}
H^{1,K_1}_{\pm}&=-\left(m^2_{B}-m^2_{K_1}\right)\Big[(C_{9}^{\text{eff}}+C_{9\ell}^{\text{NP}}+C_{9^{\prime}\ell}^{\text{NP}})
\frac{V_{1}^{K_1}}{m_{B}-m_{K_1}}+\frac{2m_{b}}{q^{2}}C_{7}^{\text{eff}}T_{2}^{K_1}\Big]\notag
\\
&\pm \sqrt{\lambda}\Big[(C_{9}^{\text{eff}}+C_{9\ell}^{\text{NP}}-C_{9^{\prime}\ell}^{\text{NP}})
\frac{A^{K_1}}{m_{B}+m_{K_1}}+\frac{2m_{b}}{q^{2}}C_{7}^{\text{eff}}T_{1}^{K_1}\Big],\notag
\\
H^{2,K_1}_{\pm}&=-(C_{10}+C_{10\ell}^{\text{NP}}+C_{10^{\prime}\ell}^{\text{NP}})\left(m_{B}+m_{K_1}\right)
V_{1}^{K_1}\pm \sqrt{\lambda}(C_{10}+C_{10\ell}^{\text{NP}}-C_{10^{\prime}\ell}^{\text{NP}})
\frac{A^{K_1}}{m_{B}+m_{K_1}},\notag
\\
H^{1,K_1}_0&=-\frac{1}{2m_{K_1}\sqrt{q^2}}\Bigg[(C_{9}^{\text{eff}}+C_{9\ell}^{\text{NP}}+C_{9^{\prime}\ell}^{\text{NP}})
\Big\{(m^2_{B}-m^2_{K_1}-q^2)\left(m_{B}+m_{K_1}\right)V_{1}^{K_1}\notag
\\
&-\frac{\lambda}{m_{B}+m_{K_1}}V_{2}^{K_1}\Big\}+2m_b C_{7}^{\text{eff}}\Big\{(m^2_{B}+3m^2_{K_1}-q^2)T_{2}^{K_1}
-\frac{\lambda}{m^2_{B}-m^2_{K_1}}T_{3}^{K_1}\Big\}
\Bigg],\notag
\\
H^{2,K_1}_0&=-\frac{1}{2m_{K_1}\sqrt{q^2}}(C_{10}+C_{10\ell}^{\text{NP}}+C_{10^{\prime}\ell}^{\text{NP}})
\Bigg[(m^2_{B}-m^2_{K_1}-q^2)\left(m_{B}+m_{K_1}\right)V_{1}^{K_1}\notag
\\
&-\frac{\lambda}{m_{B}+m_{K_1}}V_{2}^{K_1}\Bigg].\label{HA8}
\end{align}
Here $\lambda\equiv \lambda(m^2_{B}, m^2_{K_1}, q^2)$.
\subsection{Helicity amplitudes for $\Lambda_b \to \Lambda\ell^{+}\ell^{-}$ decay}
\begin{eqnarray}
H^{ij}_{nl}&=&\sum_{s_{\Lambda_b}, s_{\Lambda}}\big(\varepsilon^{\ast\alpha}(n)T^{i}_{\alpha}(s_{\Lambda_b}, s_{\Lambda})\big)\cdot\big(\overline{\varepsilon^{\ast\beta}(l)
T^{j}_{\beta}(s_{\Lambda_b}, s_{\Lambda})}\big)\notag
\\
&\equiv&\sum_{s_{\Lambda_b}, s_{\Lambda}}H^{i}_n(s_{\Lambda_b}, s_{\Lambda})\, \overline{H}^{\, j}_l(s_{\Lambda_b}, s_{\Lambda}).\label{HA9}
\end{eqnarray}
The helicity $s_{\Lambda_b}$ of the parent baryon is fixed by angular momentum conservation relation, $s_{\Lambda_b}=-s_{\Lambda}+\lambda_{j_{\text{eff}}}$.
The possible helicity configurations are shown in table \ref{table2}. Using the explicit results of the spinor matrix elements for different combinations of spin orientations, represented in appendix \ref{spinorele}, we work out the expressions of the non-vanishing helicity amplitudes
\begin{align}
H^{1}_t(\pm1/2, \mp1/2)&=\mp\frac{m_{\Lambda_b}-m_{\Lambda}}{\sqrt{q^2}}\sqrt{s_{+}}   (C_{9}^{\text{eff}}+C_{9\ell}^{\text{NP}}+C_{9^{\prime}\ell}^{\text{NP}})f^V_t\notag
\\
&-\frac{m_{\Lambda_b}+m_{\Lambda}}{\sqrt{q^2}}\sqrt{s_{-}}
(C_{9}^{\text{eff}}+C_{9\ell}^{\text{NP}}-C_{9^{\prime}\ell}^{\text{NP}})f^A_t,\notag
\\
H^{2}_t(\pm1/2, \mp1/2)&=\mp\frac{m_{\Lambda_b}-m_{\Lambda}}{\sqrt{q^2}}\sqrt{s_{+}}
(C_{10}+C_{10\ell}^{\text{NP}}+C_{10^{\prime}\ell}^{\text{NP}})f^V_t\notag
\\
&-\frac{m_{\Lambda_b}+m_{\Lambda}}{\sqrt{q^2}}\sqrt{s_{-}}
(C_{10}+C_{10\ell}^{\text{NP}}-C_{10^{\prime}\ell}^{\text{NP}})f^A_t,\notag
\end{align}
\begin{align}
H^{1}_{\pm}(\pm1/2, \pm1/2)&=\pm\sqrt{2s_-}\Big[(C_{9}^{\text{eff}}+C_{9\ell}^{\text{NP}}+C_{9^{\prime}\ell}^{\text{NP}})f^V_{\perp}
+\frac{2m_b}{q^2} C_{7}^{\text{eff}}(m_{\Lambda_b}+m_{\Lambda})f^T_{\perp}\Big]\notag
\\
&-\sqrt{2s_+}\Big[(C_{9}^{\text{eff}}+C_{9\ell}^{\text{NP}}-C_{9^{\prime}\ell}^{\text{NP}})f^A_{\perp}
+\frac{2m_b}{q^2} C_{7}^{\text{eff}}(m_{\Lambda_b}-m_{\Lambda})f^{T_5}_{\perp}\Big],\notag
\\
H^{2}_{\pm}(\pm1/2, \pm1/2)&=\pm\sqrt{2s_-}(C_{10}+C_{10\ell}^{\text{NP}}+C_{10^{\prime}\ell}^{\text{NP}})f^V_{\perp}
-\sqrt{2s_+}(C_{10}+C_{10\ell}^{\text{NP}}-C_{10^{\prime}\ell}^{\text{NP}})f^A_{\perp},\notag
\\
H^{1}_{0}(\pm1/2, \mp1/2)&=\mp\sqrt{\frac{s_-}{q^2}}\Big[(C_{9}^{\text{eff}}+C_{9\ell}^{\text{NP}}+C_{9^{\prime}\ell}^{\text{NP}})
(m_{\Lambda_b}+m_{\Lambda})f^V_{0}+2m_b C_{7}^{\text{eff}}f^T_{0}\Big]\notag
\\
&-\sqrt{\frac{s_+}{q^2}}\Big[(C_{9}^{\text{eff}}+C_{9\ell}^{\text{NP}}-C_{9^{\prime}\ell}^{\text{NP}})
(m_{\Lambda_b}-m_{\Lambda})f^A_{0}+2m_b C_{7}^{\text{eff}}f^{T_5}_{0}\Big],\notag
\\
H^{2}_{0}(\pm1/2, \mp1/2)&=\mp\sqrt{\frac{s_-}{q^2}}(C_{10}+C_{10\ell}^{\text{NP}}+C_{10^{\prime}\ell}^{\text{NP}})(m_{\Lambda_b}+m_{\Lambda})f^V_{0}\notag
\\
&-\sqrt{\frac{s_+}{q^2}}(C_{10}+C_{10\ell}^{\text{NP}}-C_{10^{\prime}\ell}^{\text{NP}})(m_{\Lambda_b}-m_{\Lambda})f^A_{0}.\label{HA10}
\end{align}
It is important here to mention that the expressions of the helicity amplitudes correspond to intermediate results and depend upon the
kinematics and polarization vectors convention. For the $\Lambda_b \to \Lambda\ell^{+}\ell^{-}$ decay, our conventions are consistent with
that used in Ref. \cite{Gutsche:2013pp}. However, the final decay observables remain same and are independent of the conventions used.
\begin{table}
\renewcommand{\arraystretch}{1}
\centering
\begin{tabular}{|c|c|c|}
\hline $\hspace{0.5cm}s_{\Lambda_b}\hspace{0.5cm}$ & $\hspace{0.5cm}s_{\Lambda}\hspace{0.5cm}$ & $\hspace{0.5cm}\lambda_{j_{\text{eff}}}\hspace{0.5cm}$ \\
\hline $+\frac{1}{2}$ & $-\frac{1}{2}$ & $0(t)$ \\
$-\frac{1}{2}$ & $+\frac{1}{2}$ & $0(t)$ \\
$+\frac{1}{2}$ & $+\frac{1}{2}$ & $1$ \\
$-\frac{1}{2}$ & $-\frac{1}{2}$ & $-1$ \\
\hline
\end{tabular}
\caption{The possible helicity configurations for $\Lambda_b \to \Lambda\ell^{+}\ell^{-}$ decay.}\label{table2}
\end{table}
\section{Formulation of physical observables}\label{physobs}
The differential decay rate in terms of helicity amplitudes for $M_{in}\to M_{f} \ell^+\ell^-$ transitions, with
$M_{in}=B, B_s$ and $M_{f}=f_0, K_0^{\ast}, K, K^{\ast}, \phi, K_1$, can be expressed as \cite{Faessler:2002ut}
\begin{eqnarray}
\frac{d\Gamma\left(M_{in}\to M_{f} \ell^+\ell^-\right)}{dq^{2}}&=&\frac{G_{F}^2\alpha^2|V_{tb}V^{\ast}_{ts}|^{2}q^2\sqrt{\lambda}\beta_l}{3.2^9 m_{in}^{3}\pi^5}\Bigg[\frac{2m_{\ell}^{2}}{q^{2}}
3\mathcal{R}e\left(H^{2,M_f}_{t}\overline{H}^{\,2,M_f}_{t}\right)\notag
\\
&+&\left(1+\frac{2m_{\ell}^{2}}{q^2}\right)\Big[H^{1,M_f^T}\overline{H}^{\,1,M_f^T}+\mathcal{R}e\left(H^{1,M_f}_{0}\overline{H}^{\,1,M_f}_{0}\right)\Big]\notag
\\
&+&\left(1-\frac{4m_{\ell}^{2}}{q^2}\right)\Big[H^{2,M_f^T}\overline{H}^{\,2,M_f^T}
+\mathcal{R}e\left(H^{2,M_f}_{0}\overline{H}^{\,2,M_f}_{0}\right)\Big]\Bigg],\label{Drateori}
\end{eqnarray}
where
\begin{eqnarray}
H^{i,M_f^T}\overline{H}^{\,i,M_f^T}\equiv \mathcal{R}e\left(H^{i,M_f}_{+}\overline{H}^{\,i,M_f}_{+}\right)
+\mathcal{R}e\left(H^{i,M_f}_{-}\overline{H}^{\,i,M_f}_{-}\right).\label{Drate1}
\end{eqnarray}
%\begin{eqnarray}
%\frac{d\Gamma\left(M_{in}\to M_{f} \ell^+\ell^-\right)}{dq^{2}}&=&\frac{G^{2}_{F}}{(2\pi)^{3}}\left(\frac{\alpha|V_{tb}V^{\ast}_{ts}|}{2\pi}\right)^{2}\frac{\sqrt{\lambda}q^{2}}{48m_{in}^{3}}
%\sqrt{1-\frac{4m_{\ell}^{2}}{q^{2}}}\Bigg[H^{1,M_f}H^{\dag1,M_f}\left(1+\frac{2m_{\ell}^{2}}{q^2}\right)\notag
%\\
%&+&H^{2,M_f}H^{\dag2,M_f}\left(1-\frac{4m_{\ell}^{2}}{q^2}\right)+\frac{2m_{\ell}^{2}}{q^{2}}
%3\mathcal{R}e\left(H^{2,M_f}_{t}H^{\dag2,M_f}_{t}\right)\Bigg],\label{Drate}
%\end{eqnarray}
%\begin{eqnarray}
%H^{i,M_f}H^{\dag i,M_f}\equiv \mathcal{R}e\left(H^{i,M_f}_{+}H^{\dag i,M_f}_{+}\right)+\mathcal{R}e\left(H^{i,M_f}_{-}H^{\dag i,M_f}_{-}\right)
%+\mathcal{R}e\left(H^{i,M_f}_{0}H^{\dag i,M_f}_{0}\right).\label{Drate112}
%\end{eqnarray}
When the final state $(M_f)$, is a vector or axial-vector, the longitudinal and transverse
polarizations can be separated and labeled as $L$ and $T$, respectively. The corresponding decay rates are written as
\begin{eqnarray}
\frac{d\Gamma(M_{in}\to M_{f}^L\ell^+\ell^-)}{dq^{2}}&=&\frac{G_{F}^2\alpha^2|V_{tb}V^{\ast}_{ts}|^{2}q^2\sqrt{\lambda}\beta_l}{3.2^9 m_{in}^{3}\pi^5}\Bigg[\frac{2m_{\ell}^{2}}{q^{2}}
3\mathcal{R}e\left(H^{2,M_f}_{t}\overline{H}^{\,2,M_f}_{t}\right)\notag
\\
&+&\left(1+\frac{2m_{\ell}^{2}}{q^2}\right)\mathcal{R}e\left(H^{1,M_f}_{0}\overline{H}^{\, 1,M_f}_{0}\right)\notag
\\
&+&\left(1-\frac{4m_{\ell}^{2}}{q^2}\right)\mathcal{R}e\left(H^{2,M_f}_{0}\overline{H}^{\,2,M_f}_{0}\right)\Bigg],\label{DrateL}
\end{eqnarray}
\begin{eqnarray}
\frac{d\Gamma(M_{in}\to M_{f}^T \ell^+\ell^-)}{dq^{2}}&=&\frac{G_{F}^2\alpha^2|V_{tb}V^{\ast}_{ts}|^{2}q^2\sqrt{\lambda}\beta_l}{3.2^9 m_{in}^{3}\pi^5}\Bigg[\left(1+\frac{2m_{\ell}^{2}}{q^2}\right)H^{1,M_f^T}\overline{H}^{\,1,M_f^T}\notag
\\
&+&\left(1-\frac{4m_{\ell}^{2}}{q^2}\right)H^{2,M_f^T}\overline{H}^{\,2,M_f^T}\Bigg],\label{DrateT}
\end{eqnarray}
Similarly, differential decay rate for $\Lambda_b \to \Lambda\ell^{+}\ell^{-}$ decay is calculated as
\begin{align}
\frac{d\Gamma\left(\Lambda_b \to \Lambda\ell^{+}\ell^{-}\right)}{dq^{2}}&=\frac{G_{F}^2\alpha^2|V_{tb}V^{\ast}_{ts}|^{2}q^2\sqrt{\lambda}\beta_l}{3.2^{10} m_{\Lambda_b}^{3}\pi^5}\Bigg[\frac{2m_{\ell}^{2}}{q^{2}}3\Big\{\left|H^{2}_{t}(+1/2, -1/2)\right|^2\notag
\\
&+\left|H^{2}_{t}(-1/2, +1/2)\right|^2\Big\}+\left(1+\frac{2m_{\ell}^{2}}{q^2}\right)
\Big\{\left|H^{1}_{+}(+1/2, +1/2)\right|^2\notag
\\
&+\left|H^{1}_{-}(-1/2, -1/2)\right|^2+\left|H^{1}_{0}(+1/2, -1/2)\right|^2
+\left|H^{1}_{0}(-1/2, +1/2)\right|^2\Big\}\notag
\\
&+\left(1-\frac{4m_{\ell}^{2}}{q^2}\right)\Big\{\left|H^{2}_{+}(+1/2, +1/2)\right|^2+\left|H^{2}_{-}(-1/2, -1/2)\right|^2\notag
\\
&+\left|H^{2}_{0}(+1/2, -1/2)\right|^2+\left|H^{2}_{0}(-1/2, +1/2)\right|^2\Big\}\Bigg].\label{Drate3}
\end{align}
The decay rate in Eq. (\ref{Drate3}) can be separated into two parts. The first part corresponding to $\Lambda_b$ and $\Lambda$ having opposite spins
is denoted as $d\Gamma\left(\Lambda_b \to \Lambda^0\ell^{+}\ell^{-}\right)/dq^2$, while the other part with $\Lambda_b$ and $\Lambda$ having same spins
is labeled as $d\Gamma\left(\Lambda_b \to \Lambda^1\ell^{+}\ell^{-}\right)/dq^2$.
The LFUV observables are constructed by taking the ratio of decay rates for $M_{in}\to M_{f}\mu^{+}\mu^{-}$
and $M_{in}\to M_{f}e^{+}e^{-}$,
\begin{eqnarray}
R_{M_{f}}\left[q^2_{\text{min}}, q^2_{\text{max}}\right]=\frac{\displaystyle\int^{q^2_{\text{max}}}_{q^2_{\text{min}}}dq^2d\Gamma(M_{in}\to M_{f}\mu^{+}\mu^{-})/dq^{2}}{\displaystyle\int^{q^2_{\text{max}}}_{q^2_{\text{min}}}dq^2d\Gamma(M_{in}\to M_{f}e^{+}e^{-})/dq^{2}}.\label{RK1}
\end{eqnarray}
For the vector and axial-vector final states, polarized LFUV ratios are defined as
\begin{eqnarray}
R_{M_{f}^{L,\,T}}\left[q^2_{\text{min}}, q^2_{\text{max}}\right]=\frac{\displaystyle\int^{q^2_{\text{max}}}_{q^2_{\text{min}}}dq^2d\Gamma(M_{in}\to
M^{L,\,T}_{f}\mu^{+}\mu^{-})/dq^{2}}{\displaystyle\int^{q^2_{\text{max}}}_{q^2_{\text{min}}}dq^2d\Gamma(M_{in}\to M^{L,\,T}_{f}e^{+}e^{-})/dq^{2}}.\label{RK2}
\end{eqnarray}
Similarly, for $\Lambda_b \to \Lambda\ell^{+}\ell^{-}$ decay
\begin{eqnarray}
R_{\Lambda}\left[q^2_{\text{min}}, q^2_{\text{max}}\right]=\frac{\displaystyle\int^{q^2_{\text{max}}}_{q^2_{\text{min}}}dq^2d\Gamma(\Lambda_b \to \Lambda\mu^{+}\mu^{-})/dq^{2}}{\displaystyle\int^{q^2_{\text{max}}}_{q^2_{\text{min}}}dq^2d\Gamma(\Lambda_b \to \Lambda e^{+}e^{-})/dq^{2}},\label{RK3}
\end{eqnarray}
\begin{eqnarray}
R_{\Lambda^{0,\,1}}\left[q^2_{\text{min}}, q^2_{\text{max}}\right]=\frac{\displaystyle\int^{q^2_{\text{max}}}_{q^2_{\text{min}}}dq^2d\Gamma(\Lambda_b \to \Lambda^{0,\,1}\mu^{+}\mu^{-})/dq^{2}}{\displaystyle\int^{q^2_{\text{max}}}_{q^2_{\text{min}}}dq^2d\Gamma(\Lambda_b \to \Lambda^{0,\,1} e^{+}e^{-})/dq^{2}}.\label{RK4}
\end{eqnarray}
\section{Predictions for LFUV ratios in the SM and the NP scenarios}\label{secNum}
\subsection{NP scenarios}
To give predictions and perform numerical analysis of the LFUV ratios, we first specify our choice of the NP scenarios, from two sets of recent global fit~\cite{Alguero:2021anc,Datta:2019zca}\footnote{Maximum likelihood fit with Gaussian distribution (or minimum $\chi^2$ fit) has been utilized in \cite{Alguero:2021anc,Datta:2019zca} by treating the theoretical and experimental covariance matrices equally. In \cite{Alguero:2021anc}, it is specified that asymmetric uncertainties have been symmetrized by taking the largest uncertainty, while in \cite{Datta:2019zca} the fit was performed with the help of MINUIT~\cite{James:1975dr}, flavio~\cite{Straub:2018kue} and Wilson~\cite{Aebischer:2018bkb} using the default configuration of these packages.} to the $b\to s\ell^+\ell^-$ data\footnote{These two sets of fit both have taken into account the measurements of LFUV ratios $R_{K^{(*)}}$, and differential branching ratios, angular observables and polarization fractions for various $b\to s\mu^+\mu^-$ channels, including $B\to K^{(*)}\mu^+\mu^-$, $B_s\to\phi\mu^+\mu^-$, $B\to X_s\mu^+\mu^-$ and $B_s\to\mu^+\mu^-$, measured by different collaborations including LHCb, CMS, ATLAS, Belle, BaBar and CDF. In the more recent analysis~\cite{Alguero:2021anc}, further updates on $R_K$~\cite{LHCb:2021trn}, branching ratios $\mathcal B(B^{0,+}\to K^{0,+}\mu^+\mu^-)$~\cite{BELLE:2019xld} and $\mathcal B(B_s\to \mu^+\mu^-)$~\cite{Ferreres-Sole:2021qxv}, and angular distribution of $B^+\to K^{(*)+}\mu^+\mu^-$~\cite{LHCb:2020gog,CMS:2018qih} and $B\to K^*e^+e^-$~\cite{LHCb:2020dof} were also included. For details, see \cite{Alguero:2021anc,Datta:2019zca} and the references therein. For the calculation of observables,  $B\to K^{(*)}$ form factors in \cite{Khodjamirian:2010vf} (LCSR) and \cite{Gubernari:2018wyi} (LCSR+lattice QCD) were respectively adopted in \cite{Alguero:2021anc} and \cite{Datta:2019zca} (using flavio), and $B_s\to \phi$ form factors in \cite{Straub:2015ica} (LCSR+lattice QCD) were used in both \cite{Alguero:2021anc} and \cite{Datta:2019zca}. Besides, the charm loop effects identified in \cite{Khodjamirian:2010vf} were also considered in \cite{Alguero:2021anc,Datta:2019zca}.}, which could be easily realized in the specific simple NP models.
\begin{enumerate}[label=\arabic*)]
	\item Assuming LFUV NP in $b\to s\mu^+\mu^-$ only, two basic (1D) NP scenarios (S1) $C_{9\mu}^{\text{NP}}$ and (S2) $C_{9\mu}^{\text{NP}}=-C_{10\mu}^{\text{NP}}$, continue to provide better fit to all data, including the latest experimental inputs \cite{Alguero:2021anc,Datta:2019zca}. Therefore, for the S1 and S2 scenarios, we consider the best-fit Wilson coefficients from table-1 of Ref. \cite{Alguero:2021anc}, and collect them in table \ref{tab:bestfitWC}, for the sake of completeness. S1 and S2 can be realized in the simplest
NP models involving the tree-level exchange of a leptoquark (LQ) or a $Z^{\prime}$ boson. While S1 is only possible with a $Z^{\prime}$, S2 can appear
in both LQ and $Z^{\prime}$ models \cite{Alok:2017sui}.
\begin{table*}[!htbp]
\renewcommand{\arraystretch}{1}
	\begin{center}
			\begin{tabular}{|clcc|}
				\hline
				Scenario & & Best-fit value & $1\sigma$ \\
				\hline
                S1 & $C_{9\mu}^{\text{NP}}$                          &$-1.06$       & $[-1.20, -0.91]$  \\
                S2 & $C_{9\mu}^{\text{NP}}=-C_{10\mu}^{\text{NP}}$   &$-0.44$       & $[-0.52, -0.37]$  \\
                S3 & $C_{9\mu}^{\text{NP}}=-C_{10\mu}^{\text{NP}}$   &$-0.67$       & $[-0.82, -0.52]$  \\
                   & $C_{9e}^{\text{NP}}=-C_{10e}^{\text{NP}}$       &$-0.28$       & $[-0.48, -0.08]$  \\
                S4 & $C_{9\mu}^{\text{NP}}=-C_{10\mu}^{\text{NP}}$   &$-0.64$       & $[-0.78, -0.50]$  \\
                   & $C_{9e}^{\text{NP}}$                            &$-0.65$       & $[-1.09, -0.21]$  \\
                S5 & $C_{9\mu}^{\text{V}}=-C_{10\mu}^{\text{V}}$     &$-0.30$       & $[-0.39, -0.21]$  \\
                   & $C_{9}^{\text{U}}$                              &$-0.92$       & $[-1.10, -0.72]$  \\
                S6 & $C_{9\mu}^{\text{V}}$                           &$-1.12$       & $[-1.28, -0.95]$  \\
                   & $C_{10^{\prime}}^{\text{U}}$                    &$-0.31$       & $[-0.46, -0.15]$  \\
                \hline
	\end{tabular}		
			\caption{Best-fit values of the Wilson coefficients, and the $1\sigma$ ranges of different NP scenarios with assumptions, such as, purely LFUV NP in $b\to s\mu^+\mu^-$, additional arbitrary LFUV NP in $b\to s e^+e^-$ along with LFUV NP in $b\to s\mu^+\mu^-$, and both LFU and
LFUV NP.}\label{tab:bestfitWC}
	\end{center}
\end{table*}
	\item  Motivated by removing the tensions between the separate fits to LFD and LFUV observables, we consider the NP scenarios extending S1 and S2, with additional arbitrary LFUV NP in $b\to se^+e^-$, which affects only LFUV observables, leading to improved pulls with respect to the SM. While several
scenarios extending S1 and S2, with the addition of one nonzero NP WC in $b\to se^+e^-$ are reported in \cite{Datta:2019zca}, we pick only those scenarios,
which can be realized in the context of the LQ and $Z^{\prime}$ models, and have improved pulls with respect to the SM, compared to the ones obtained in S1 and S2. Therefore, we consider S3 and S4 from table-4 of Ref. \cite{Datta:2019zca}, that can be generated in $Z^{\prime}$ model, whereas only S3 can be realized in the LQ models due to the fact that leptoquarks can only contribute to $C_{9\ell}^{\text{NP}}=-C_{10\ell}^{\text{NP}}$, $\ell=e, \mu$. S3 and S4 are listed in table \ref{tab:bestfitWC}.
    \item Next, we consider the NP hypothesis which allows LFU NP (equal contributions for all the lepton flavours), in addition to LFUV contributions to
muons only. NP Wilson coefficients in this case can be represented as
\begin{eqnarray}
C_{i^{(\prime)}e}^{\text{NP}}=C_{i^{(\prime)}}^{\text{U}},\qquad\quad C_{i^{(\prime)}\mu}^{\text{NP}}=C_{i^{(\prime)}}^{\text{U}}
+C_{i^{(\prime)}\mu}^{\text{V}},\label{WC910}
\end{eqnarray}
with $i=9, 10$, for the $b\to s\mu^+\mu^-$ and $b\to se^+e^-$ transitions, respectively. The superscript ``U'' and ``V'' represents the
LFU and LFUV contribution, respectively. Several NP scenarios
with both LFU and LFUV NP contributions are presented in table-4 of Ref. \cite{Alguero:2021anc}. For the sake of simplicity,
we restrict to the NP scenarios, which only extend S1 and S2, yielding equal or improved pulls compared to the corresponding ones for the S1 and S2 scenarios, given in table-1 of \cite{Alguero:2021anc}, and can be fairly easily realized in specific NP models. It is important to mention that one
should be very careful while comparing pulls found in different analyses, as they strongly depend on the choice of observables, treatment of the
theoretical errors, and the fact that how the analysis is performed. Therefore, we only consider comparison of pulls between scenarios
obtained within a single analysis. Based on the above criteria, we consider S8 and S11 given in table-4 of \cite{Alguero:2021anc}, and label them
as S5 and S6, as shown in table \ref{tab:bestfitWC}. Scenario S5 can be generated via off-shell
photon penguins \cite{Crivellin:2018yvo} in a LQ model, while S6 can be generated in $Z^{\prime}$ model with vector couplings
to muons and additional Vectorlike quarks with the quantum numbers of left-handed quarks doublets \cite{Bobeth:2016llm}.
\end{enumerate}
It is worth mentioning that the above two methods of considering additional $b\to se^+e^-$ NP are complementary and each NP scenario
in one method can be translated into the other, and vice versa \cite{Kumar:2019qbv}, however, they offer distinct fitting mechanism to
LFD and LFUV observables and therefore may correspond to unique NP predictions. For example, it is suggested in Ref. \cite{Alguero:2018nvb},
that assuming both LFU and LFUV NP provides a different mechanism to obey the constraint from the LFD observable $\mathcal{B}(B_s\to\mu^+\mu^-)$, with large
value of $C_{10\mu}^{\text{V}}$ WC with opposite sign $C_{10}^{\text{U}}$ WC value, and hence allows the possibility of new class of NP models
with large LFU and LFUV contributions to $C_{10\mu}$ at the same time, to account for the combined LFD+LFUV observables. This result is not obtained with only LFUV NP contributions to both $b\to s\mu^+\mu^-$ and $b\to se^+e^-$, as the additional LFUV NP in $b\to se^+e^-$,
affects only LFUV observables and the LFD observables, in this case, explained only by the LFUV NP contributions to $b\to s\mu^+\mu^-$, lead to the other
favoured NP scenarios with large pulls.
\subsection{Predictions for the LFUV ratios}
In this section, we explore the patterns of the lepton flavour universality violation, in different bins of the complementary ratios, due the presence of different NP possibilities in the form of the best-fit values of the Wilson coefficients, found in the recent global fit analyses. For that, we consider various LFUV ratios, including (pseudo-)scalar final states, $R_{f_0}$, $R_{K^*_0}$, $R_{K}$,
unpolarized and polarized (axial-)vector final states, $R_{\phi^{(L,\,T)}}$, $R_{{K^*}^{(L,\,T)}}$, $R_{K_1^{(L,\,T)}(1270,\,1400)}$, and
$\Lambda$ baryon final state with different spin orientations $R_{\Lambda^{(0,\,1)}}$. Experimentally, $R_{K^{(*)}}$ has already been measured by LHCb in the kinematical region $q^2\leq 6$ GeV$^2$, and by Belle in the low and high $q^2$ regions with large errors. Future precision measurements of high $q^2$ bins at Belle II and LHCb will be complementary and important for testing LFU, therefore, in our analysis, we consider only high $q^2$ bins of $R_{K^{(*)}}$.

In figures~\ref{Fig:SPSPlot}-\ref{Fig:RLamda}, we show the SM and the NP predictions, for the LFUV ratios in the low $q^2$ bin, [0.045, 1] GeV$^2$, the central $q^2$ bin, [1, 6] GeV$^2$, and the high $q^2$ bin, [14, $q^2_{\text{max}}$] GeV$^2$. The height of the bars in figures~\ref{Fig:SPSPlot}-\ref{Fig:RLamda}, represent the uncertainty in the SM and NP predictions due to the errors in the form factors. We explicitly list the central values predictions and the predictions due to the errors in the form factors, for all the LFUV ratios in the SM and the NP scenarios, in tables~\ref{tab:Rf0bin}-\ref{tab:RLambdabin}, of appendix~\ref{app:LFUV}.
%Furthermore, we also give the deviations of LFUV ratios due to the $1\sigma$ ranges of the best-fit Wilson coefficients in these tables.
\subsubsection{SM and NP predictions for \texorpdfstring{$R_{S}$}{} and \texorpdfstring{$R_{P}$}{}}
\begin{figure*}[!t]
\centering
%\begin{center}
\includegraphics[scale=0.55]{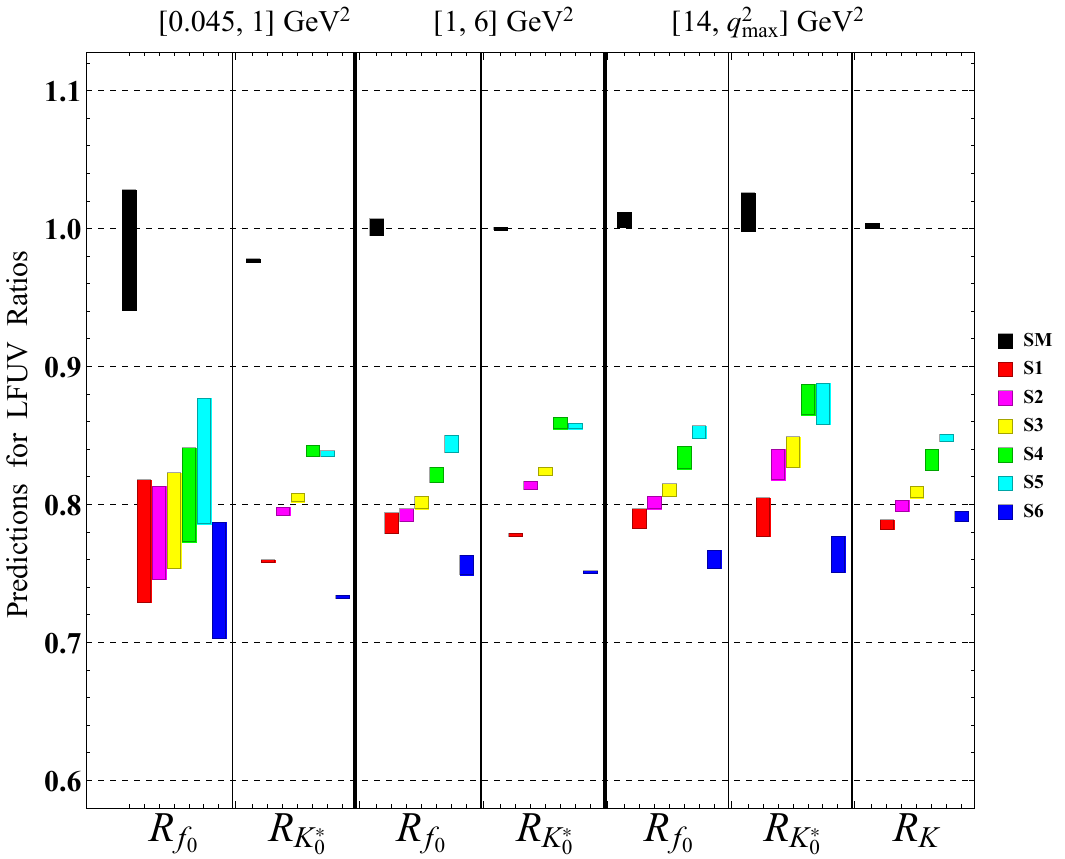}
%\end{center}
\caption{Predictions for the LFUV ratios involving decays with scalar or pseudoscalar final state particles, $R_{f_0}$, $R_{K^*_0}$, and $R_{K}$. Three kinematical regions, low $[0.045, 1]$ GeV$^2$, central $[1, 6]$ GeV$^2$, and high $[14, q^2_{\text{max}}]$ GeV$^2$, are chosen, where
$q^2_{\text{max}}= 19.2, 14.9$, and $22.9$ GeV$^2$, for $R_{f_0}$, $R_{K^*_0}$, and $R_{K}$, respectively. In each case, predictions from left to right,
correspond to the SM and scenarios S1 to S6, depicted with different colors.}
\label{Fig:SPSPlot}
\end{figure*}
In this section, we consider the LFUV ratios involving decays with scalar or pseudoscalar final states, $R_{f_0}$, $R_{K^*_0}$, and $R_{K}$. SM and NP predictions for these ratios are shown in figure~\ref{Fig:SPSPlot}. It is clear from figure~\ref{Fig:SPSPlot}, that in all $q^2$ bins, NP predictions for
these ratios are considerably lower than the corresponding SM predictions. Considering, $R_{f_0}$ first, the SM predictions of $R_{f_0}$ in the central and high $q^2$ bins are relatively clean and these bins are also very useful to distinguish among the
different NP scenarios except between S1 and S2 scenarios. This means that the NP sensitivities vary for the different scenarios with the highest NP sensitivity observed in scenario S6.
Next, for $R_{K^*_0}$, SM values in the low and central $q^2$ bins are very clean therefore future measurements of $R_{K^*_0}$ in these bins have
the potential to reveal NP unambiguously, however in order to differentiate the NP scenarios very precise measurements of $R_{K^*_0}$
will be required in these two bins as the form factor uncertainties in the presence of the NP scenarios also largely cancel out. On the other hand, the SM predictions in the low $q^2$ bin of $R_{f_0}$, and the high $q^2$ bin of $R_{K^*_0}$ are not clean and also the NP predictions involve large uncertainties, resulting in the frequent overlap of the different NP scenarios, making such bins less useful. In any case, NP predictions for $R_{f_0}$ and $R_{K^*_0}$ should differ from the SM predictions, therefore it would be very useful for testing LFU by measuring them. In addition, very interestingly, the measurement of $R_K$ at high $q^2$, which can be accessible at Belle II \cite{Kou:2018nap}, can help to distinguish almost all NP scenarios, making such measurement very anticipated.
\subsubsection{SM and NP predictions for \texorpdfstring{$R_{V^{(L,\,T)}}$}{} and \texorpdfstring{$R_{A^{(L,\,T)}}$}{}}
\begin{figure*}[b]
\centering
%\begin{center}
\includegraphics[scale=0.50]{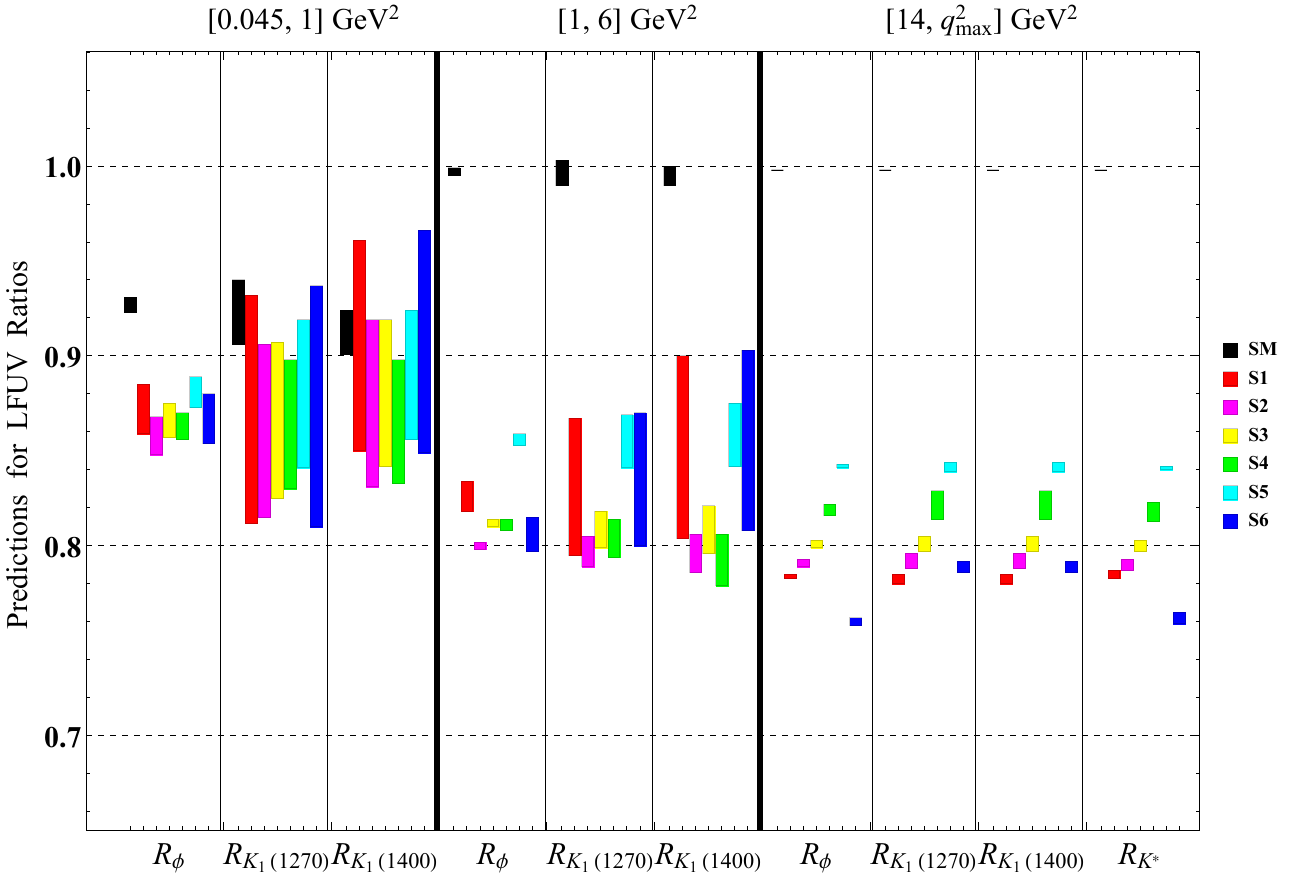}
%\end{center}
\caption{Predictions for the LFUV ratios involving decays with vector or axial-vector final state particles, $R_{\phi}$, $R_{K^*}$, $R_{K_1(1270)}$, and $R_{K_1(1400)}$, where only $R_{K_1(1270)}$ values with $\theta_{K_1}=-34^{\circ}$, and $R_{K_1(1400)}$ values with $\theta_{K_1}=34^{\circ}$ are presented.
Three kinematical regions, low $[0.045, 1]$ GeV$^2$, central $[1, 6]$ GeV$^2$, and high $[14, q^2_{\text{max}}]$ GeV$^2$, are chosen, where
$q^2_{\text{max}}= 18.9, 19.2, 16$, and $15$ GeV$^2$, for $R_{\phi}$, $R_{K^*}$, $R_{K_1(1270)}$, and $R_{K_1(1400)}$ respectively. In each case, predictions from left to right, correspond to the SM and scenarios S1 to S6, depicted with different colors.}
\label{Fig:VAPlot}
\end{figure*}
In this section, we consider the LFUV ratios involving decays with unpolarized and polarized vector or axial-vector final states,
$R_{\phi^{(L,\,T)}}$, $R_{{K^*}^{(L,\,T)}}$, $R_{K_1^{(L,\,T)}(1270)}$, and $R_{K_1^{(L,\,T)}(1400)}$. Before presenting our predictions, we need to specify what value of the $K_1$ mixing angle $\theta_{K_1}$, we adopt. In fact, there are two widely used values, i.e., $\theta_{K_1}\sim-34^{\circ}$ \cite{Hatanaka:2008xj}, from $B\to K_1\gamma$ and $\tau\to K_1(1270)\nu$, and $\theta_{K_1}\sim34^{\circ}$ \cite{Cheng:2011pb,Tanabashi:2018oca}, from the study of the $f_1(1285)-f_1(1420)$ and $h_1(1170)-h_1(1380)$ mixing~\footnote{It has been proposed in~\cite{Hayasaka:2021ecj} to extract the $K_1$ mixing angle from $\tau^-\to K_1^-\nu_\tau \to (K^-\omega) \nu_\tau\to (K^- \pi^+\pi^-\pi^0)\nu_\tau \ $.}. These different possibilities of $\theta_{K_1}$ lead to different predictions for the observables. In the case of $\theta_{K_1}=-34^{\circ}$, the branching ratio of $\mathcal B(B\to K_1(1400)\ell^+\ell^-)$ is suppressed by one to two orders of magnitude with respect to $\mathcal B(B\to K_1(1270)\ell^+\ell^-)$, while in the case of $\theta_{K_1}=34^{\circ}$, the situation is reversed,
such that $\mathcal B(B\to K_1(1270)\ell^+\ell^-)$ is more suppressed. Given the highly suppressed decay modes are difficult to measure experimentally, we only present the enhanced mode for each possibility of $\theta_{K_1}$, i.e., $B\to K_1(1270)\ell^+\ell^-$, for $\theta=-34^\circ$, and $B\to K_1(1400)\ell^+\ell^-$, for $\theta=34^\circ$. In fact, these two cases have very analogous predictions for $R_{K_1}$ as can be seen in the subsequent analysis.
\begin{figure*}[b]
\centering
%\begin{center}
\includegraphics[scale=0.50]{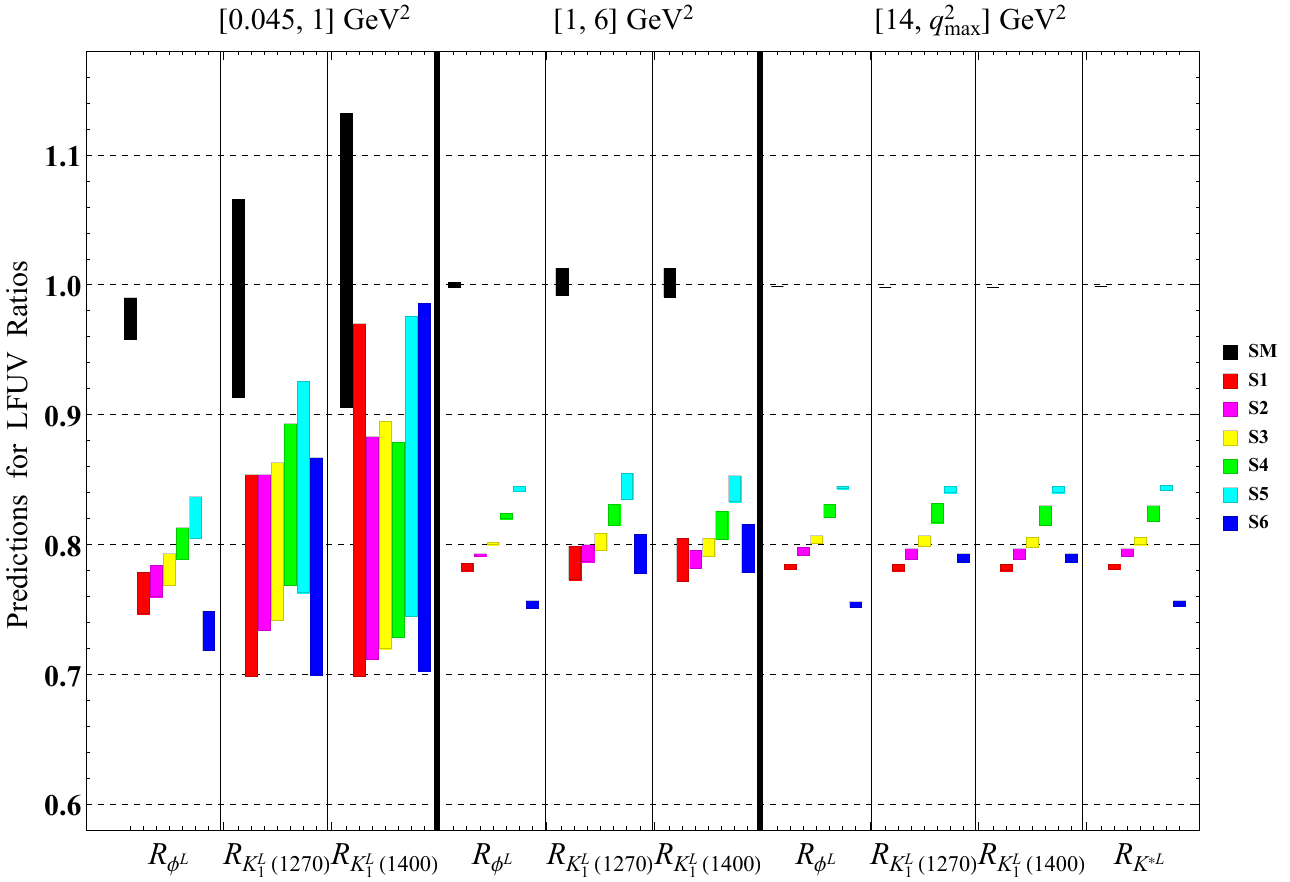}
%\end{center}
\caption{Same as figure \ref{Fig:VAPlot}, where final state particles are longitudinally polarized, giving polarized LFUV ratios,
$R_{\phi^L}$, $R_{{K^*}^L}$, $R_{K_1^L(1270)}$, and $R_{K_1^L(1400)}$.}
\label{Fig:VALPlot}
\end{figure*}

In figure~\ref{Fig:VAPlot}, we present the SM and the NP predictions for the unpolarized LFUV ratios, $R_{\phi}$, $R_{{K^*}}$, and $R_{K_1(1270,\,1400)}$.
SM predictions of these LFUV ratios in the central and high $q^2$ region are close to one, while in the low $q^2$ region [0.045, 1] GeV$^2$, due to non-negligible lepton mass effects \cite{Hiller:2003js}, they are less than one. Further, considering the SM predictions, in the low $q^2$
bin [0.045, 1] GeV$^2$, it is important to mention that the involved branching fractions in these LFUV ratios are dominated by $C_{7^{(\prime)}}$, instead of $C_{9,10}$, and these magnetic dipole Wilson coefficients enter in the helicity amplitudes corresponding to the vector leptonic current. However, compared to $C_7$, $C_{7^{\prime}}$ is still $m_s/m_b$ suppressed, and therefore we have ignored it for the SM predictions of these unpolarized LFUV ratios. Furthermore, for the NP predictions, in the low $q^2$ region, although the branching fractions are highly sensitive to NP scenarios $C_{7^{(\prime)}}^{\text{NP}}$, contributions of these radiative coefficients $C_{7^{(\prime)}}^{\text{NP}}$ to the LFUV ratios, involving both muons and electrons, are lepton flavor universal, and therefore they can only play a subleading role in interference terms involving additional semileptonic NP coefficients \cite{Capdevila:2017bsm}. So, in our study, we do not further consider scenarios with $C_{7^{(\prime)}}^{\text{NP}}$, and for the NP predictions of these unpolarized LFUV ratios, in the
low $q^2$ region, we use the already selected NP scenarios.

In figure~\ref{Fig:VAPlot}, we observe that in the low $q^2$ region [0.045, 1] GeV$^2$, $R_\phi$ is able to discriminate between the SM and the NP values although it cannot distinguish any specific NP scenario, and on the contrary $R_{K_1(1270,\,1400)}$, in the same $q^2$ bin, do not have good sensitivity to NP as the NP predictions overlap with the SM ranges, which also have relatively large uncertainties. With the increase of the momentum transfer, $R_\phi$ and $R_{K_1}$ become more sensitive to NP, and in the central $q^2$ region [1, 6] GeV$^2$, $R_\phi$ can distinguish S1 and S5, while $R_{K_1(1270,\,1400)}$ are not able to discriminate any specific NP scenario. The measurement of $R_\phi$ in this region is very useful given the statistical uncertainty can be less than 0.05 after 50 fb$^{-1}$ data is accumulated at LHCb~\cite{Bediaga:2018lhg}. Furthermore, in the high $q^2$ region, sensitivity to NP becomes even more clear as both $R_\phi$ and $R_{K_1(1270,\,1400)}$ have very small errors for the SM and NP predictions, and thus should be able to distinguish among most NP scenarios except between S2 and S6 in case of $R_{K_1(1270,\,1400)}$. Besides, the high $q^2$ bin of $R_{K^*}$ is also very useful for differentiating among the NP scenarios, therefore future measurements of both $R_K$ and $R_{K^*}$ in high $q^2$ region would be crucial for probing NP signature in the form of LFUV, given the Belle II sensitivities are less than $4\%$ with 50 ab$^{-1}$ data~\cite{Kou:2018nap}.
\begin{figure*}[b]
\centering
%\begin{center}
\includegraphics[scale=0.50]{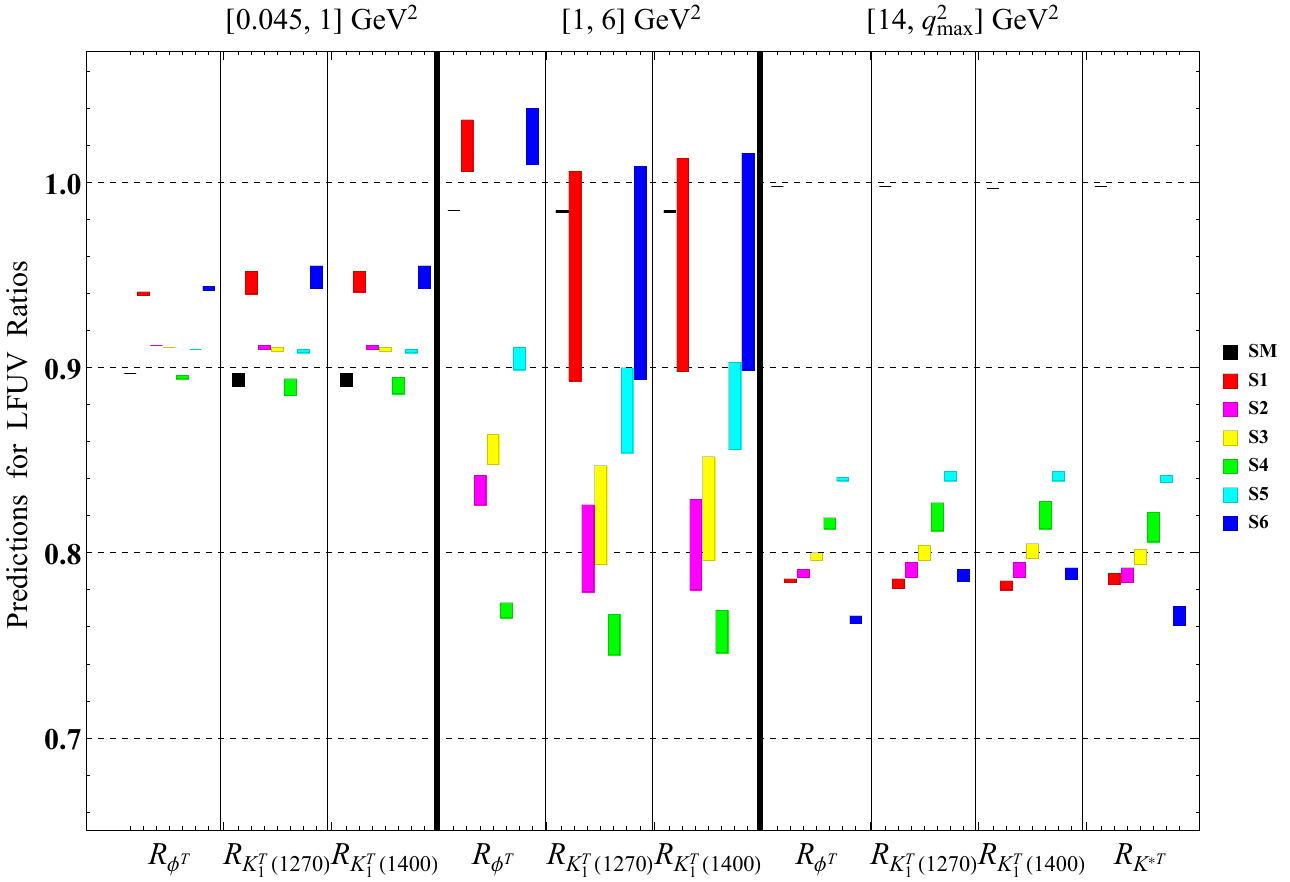}
%\end{center}
\caption{Same as figure \ref{Fig:VAPlot}, where final state particles are transversely polarized, giving polarized LFUV ratios,
$R_{\phi^T}$, $R_{{K^*}^T}$, $R_{K_1^T(1270)}$, and $R_{K_1^T(1400)}$.}
\label{Fig:VATPlot}
\end{figure*}

Additionally, in figure~\ref{Fig:VALPlot}, we show the results for the LFUV ratios in the presence of NP scenarios, with the final vector and axial-vector states longitudinally polarized, $R_{\phi^L}$, $R_{K_1^L(1270,\,1400)}$ and $R_{{K^*}^L}$. These ratios can provide complementary information for testing the
LFU. Although the ratios corresponding to longitudinally polarized final states, i.e., $R_{\phi^L}$, $R_{K_1^L(1270,\,1400)}$ and $R_{{K^*}^L}$ have similar behaviours with respect to $R_{\phi}$, $R_{K_1(1270,\,1400)}$ and $R_{K^*}$, $R_{\phi^L}$ and $R_{K_1^L}$ are more sensitive to NP in the central $q^2$ region, with $R_{\phi^L}$ giving very distinct NP predictions for almost all the NP scenarios. In contrast, the LFUV ratios for transversely polarized final state mesons, as shown in figure~\ref{Fig:VATPlot}, have even more interesting behaviours in the low $q^2$ region: they are sensitive to effects from the NP scenarios except S4 because in these scenarios they are greater than the SM predictions and with small errors. In the central $q^2$ region, $R_{\phi^T}$ in different NP scenarios except S1 and S6 is quite distinct and clearly distinguishable from the clean SM prediction with NP scenarios showing sensitivity to both the positive and negative side of the SM, while in the same $q^2$ bin $R_{K_1^{T}(1270,\,1400)}$ in different NP scenarios have relatively large errors, making it hard to discriminate among the NP scenarios except S4. In the high $q^2$ region, analogous to $R_{\phi^{L}}$, $R_{K_1^{L}(1270,\,1400)}$ and $R_{{K^*}^L}$, the ratios for transverse polarization $R_{\phi^{T}}$, $R_{K_1^{T}(1270,\,1400)}$ and $R_{{K^*}^T}$ in both the SM and the NP scenarios have relatively small errors and they provide very good sensitivity to test the lepton flavor universality.
\subsubsection{SM and NP predictions for \texorpdfstring{$R_{\Lambda^{(0,\,1)}}$}{}}
\begin{figure*}[!t]
\centering
%\begin{center}
\includegraphics[scale=0.55]{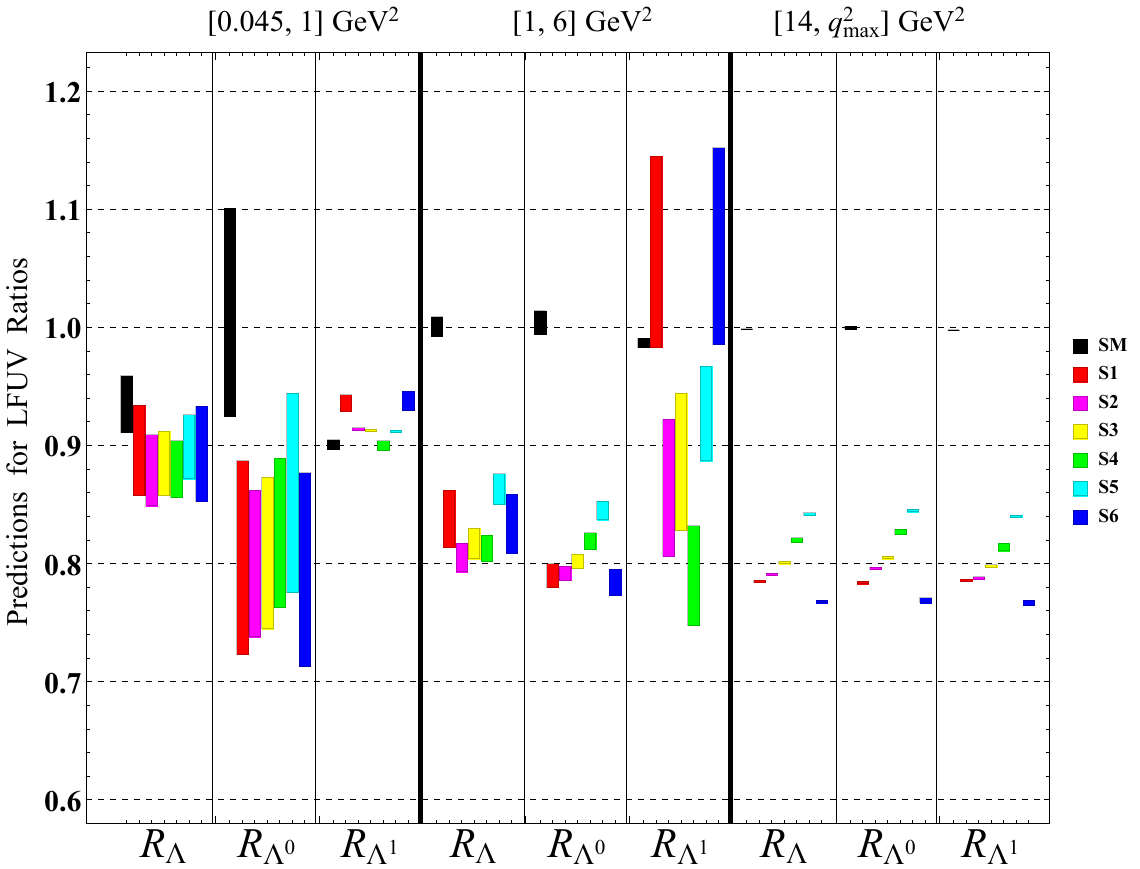}
\caption{Predictions for the LFUV ratios, $R_{\Lambda}$, $R_{\Lambda^0}$, and $R_{\Lambda^1}$, involving baryonic final state. Three kinematical regions, low $[0.045, 1]$ GeV$^2$, central $[1, 6]$ GeV$^2$, and high $[14, q^2_{\text{max}}]$ GeV$^2$, are chosen, where
$q^2_{\text{max}}= 20.3$ GeV$^2$. In each case, predictions from left to right, correspond to the SM and scenarios S1 to S6, depicted with different colors.}
\label{Fig:RLamda}
%\end{center}
\end{figure*}
In this section, we consider the LFUV ratios $R_{\Lambda^{(0,\,1)}}$. SM and NP predictions for the scenarios S1-S6, for the ratios $R_{\Lambda}$, $R_{\Lambda^0}$, and $R_{\Lambda^1}$ are presented in figure~\ref{Fig:RLamda}. It is observed that the behaviour of $R_\Lambda$ and $R_{\Lambda^{0}}$ is analogous in the sense that the large-recoil bins having low sensitivity to NP cannot distinguish among the different NP scenarios, the central $q^2$ bins with increased sensitivity to NP scenarios can partially distinguish the NP scenarios, and the low-recoil bins with distinct and clean SM and NP predictions can well distinguish all the NP scenarios. In contrast, for $R_{\Lambda^1}$, most NP scenarios are non-distinguishable by using the central $q^2$ bins, partially distinguishable by using the high-recoil bins and almost fully distinguishable by using the low-recoil bins. Therefore, the most remarkable conclusion on $R_{\Lambda^{(0,\,1)}}$ is that it would be most helpful to measure the high $q^2$ bins of $R_{\Lambda^{(0,\,1)}}$ because these bins have very small uncertainties. Lastly, similar to $R_{\phi^T}$, $R_{\Lambda^1}$ corresponding to S1 and S6, in central $q^2$ region may exceed 1, which can be an interesting characteristic for these scenarios, although they are not distinguishable from each other.
\section{Summary and conclusions}\label{summary}
In recent years, a number of experimental measurements for the $b\to s\ell^+\ell^-$ transitions have shown deviations from the SM expectations. Such measurements include the branching ratios $\mathcal B(B\to K^{(*)}\mu^+\mu^-)$ and $\mathcal B(B_s\to \phi\mu^+\mu^-)$, the angular observables in $B\to K^*\mu^+\mu^-$ decay including the famous $P_5^{\prime}$ anomaly, and very importantly, the LFUV ratios $R_{K^{(*)}}$ which are ``clean'' probe for LFUV/NP. On the other hand, experimental measurements of the LFUV ratios for either more $b\to s\ell^+\ell^-$ channels or more kinematical regions at Belle II and LHCb have been put on the agenda~\cite{Kou:2018nap,Bediaga:2018lhg}. In light of the current stage, we have studied the LFUV ratios for various $b\to s\ell^+\ell^-$ channels with (pseudo-)scalar and (axial-)vector final state mesons including $R_{f_0}$, $R_{K_0^{\ast}}$, $R_K$, $R_{K^{\ast}}$, $R_\phi$, $R_{K_1}$ as well as  $R_\Lambda$ for $\Lambda_b\to\Lambda\ell^+\ell^-$. In particular, for the cases when spin-1 meson or the $\Lambda$ baryon is the final state, we have also considered the LFUV ratios with the final state hadron longitudinally and transversally polarized.

In our calculation, we have adopted the recent results of hadronic form factors calculated in lattice QCD or/and QCD light-cone sum rules (LCSR). Within the framework of the effective field theory, we have studied various decay channels by employing the helicity formalism, and give the expressions of the physical observables in terms of the helicity amplitudes by keeping lepton mass effects. Further, we have explicitly worked out
the expressions of the helicity amplitudes in terms of the (NP) Wilson coefficients and the general hadronic form factors, in a consistent
manner, by using the same kinematical configuration and polarization conventions for all the decay channels, which allows others
to easily check and use our expressions.

In the numerical analysis, we have made predictions and performed analysis for the SM and the selected NP scenarios. Given
the updated measurements of $R_{K^{(*)}}$ suggest NP also present in $b\to se^+e^-$, therefore, besides considering
the two basic $b\to s\mu^+\mu^-$ NP scenarios S1 ($C_{9\mu}^{\text{NP}}$ only) or S2 ($C_{9\mu}^{\text{NP}}=-C_{10\mu}^{\text{NP}}$),
we have also considered four NP scenarios which extend S1 and S2, assuming additional arbitrary LFUV NP in $b\to se^+e^-$ or both LFU and LFUV NP,
and have explicit model interpretations. These scenarios are two $b\to s\mu^+\mu^-$ plus $b\to se^+e^-$  NP scenarios S3 and S4, and two LFU plus LFUV NP scenarios S5 and S6. The conclusions can be summarized as follows:
\begin{itemize}
  \item $R_{K^{(*)}}$ in the high $q^2$ region have quite good sensitivity to NP, therefore future precision measurements on the high $q^2$ bins will be important complement to the measurements of low $q^2$ bins in probing LFUV/NP in the $b\to s\ell^+\ell^-$ transition.
  \item Measurements of the LFUV ratios with scalar mesons in final states are also very helpful for distinguishing NP scenarios. In particular, $R_{f_0}$ in the bins [1, 6] GeV$^2$ and $[14,\,q^2_{\text{max}}]$ GeV$^2$, and $R_{K^{*}_0}$ in the bins [0.045, 1] GeV$^2$ and [1, 6] GeV$^2$ are useful because the theoretical uncertainties in these bins are relatively small compared with other bins.
  \item $R_\phi$ is useful for testing LFUV/NP in all kinematical regions and especially in the high $q^2$ region where theoretical predictions have small errors and the different NP scenarios have distinct predictions. In contrast, $R_{K_1}$ in the SM and NP scenarios have larger errors
       in the low and central $q^2$ bins and it has good sensitivity to NP in the central and high $q^2$ region, with NP scenarios more distinct
       in the high $q^2$ region.
  \item $R_{\phi^L}$ and $R_{K_1^L}$ corresponding to longitudinally polarized final state meson have similar behaviours with respect to the unpolarized ratios $R_\phi$ and $R_{K_1}$ in all $q^2$ bins, but for $R_{\phi^T}$ and $R_{K_1^T}$ corresponding to transversely polarized $\phi$ and $K_1$, the low and high $q^2$ bins have less uncertainties in the NP predictions, while the central $q^2$ bins NP predictions have large uncertainties.
  \item $R_{\Lambda}$, $R_{\Lambda^0}$ and $R_{\Lambda^1}$ are all very sensitive to NP with tiny theoretical errors in high $q^2$ region, which can test LFUV with distinct NP predictions. $R_{\Lambda^{(0,\,1)}}$, in the region [0.045, 1] GeV$^2$ do not have good sensitivity to NP, while in [1, 6] GeV$^2$, the measurements of $R_{\Lambda^{0}}$ can distinguish some of the NP scenarios, e.g., the central $q^2$ bin of $R_{\Lambda^{0}}$ can distinguish S4 and S5.
\end{itemize}
In conclusion, similar to $R_{K^{(*)}}$, SM predictions for the various complementary LFUV ratios are
theoretically clean in different kinematical regions and have high sensitivity to NP. Therefore, the future measurements on the LFUV ratios for these additional channels, along with the more precise $R_{K^{(*)}}$ measurements, can provide critical information on testing NP/LFUV in the $b\to s\ell^+\ell^-$ FCNC transitions. In addition, LFUV ratios with polarized final state particles are also found to be sensitive to different NP scenarios, and therefore can provide additional complementary probe of LFUV. With the running of Belle II and future upgrade of LHCb,
the measurements of many LFUV ratios studied in this work will be accessible, especially $R_{K^{(*)}}$ in high $q^2$ region
and $R_\phi$ which have already been planned~\cite{Kou:2018nap,Bediaga:2018lhg}. We hope upcoming experimental and theoretical
studies on the LFUV ratios in the $b\to s\ell^+\ell^-$ transitions, along with giving crucial evidence of possible NP behind
the $b\to s\ell^+\ell^-$ anomalies, will also help to identify the true structure of the underlying NP, by differentiating among
the emerging NP scenarios.
\section*{Acknowledgments}
F.M.B would like to acknowledge the financial support from the provincial postdoctoral talent introduction fund under grant No. 2019YJ-01. The work is partly supported by National Science Foundation of China under the Grants 11775012, 11521505 and 11621131001. Z.R.H would like to acknowledge the YST Program at the APCTP. M.A.P would like to thank the hospitality provided by IHEP when this project was initiated. We would like to thank Yixiong Zhou for useful discussions.
\appendix
\section{SM Wilson coefficients}\label{append}
For the explicit form of the Wilson coefficients $C_{7}^{\text{eff}}(q^{2})$ and $C_{9}^{\text{eff}}(q^{2})$, we follow \cite{Bobeth:1999mk,Beneke:2001at,Asatrian:2001de,Asatryan:2001zw,Greub:2008cy,Du:2015tda}.
For the sake of completeness we give the expressions of these Wilson coefficients used in our study
\begin{eqnarray}
C_{7}^{\text{eff}}(q^2)&=&C_{7}-\frac{1}{3}\left(C_{3}+\frac{4}{3}C_{4}+20C_{5}+\frac{80}{3}C_{6}\right)
-\frac{\alpha_{s}}{4\pi}\left[(C_{1}-6C_{2})F^{(7)}_{1,c}(q^2)+C_{8}F^{(7)}_{8}(q^2)\right],\notag\\
C_{9}^{\text{eff}}(q^2)&=&C_{9}+\frac{4}{3}\left(C_{3}+\frac{16}{3}C_{5}+\frac{16}{9}C_{6}\right)
-h(0, q^2)\left(\frac{1}{2}C_{3}+\frac{2}{3}C_{4}+8C_{5}+\frac{32}{3}C_{6}\right)\notag\\
&-&h(m_{b}^{\text{pole}}, q^2)\big(\frac{7}{2}C_{3}+\frac{2}{3}C_{4}+38C_{5}+\frac{32}{3}C_{6}\big)+h(m_{c}^{\text{pole}}, q^2)
\big(\frac{4}{3}C_{1}+C_{2}+6C_{3}+60C_{5}\big)\notag\\
&-&\frac{\alpha_{s}}{4\pi}\left[C_{1}F^{(9)}_{1,c}(q^2)+C_{2}F^{(9)}_{2,c}(q^2)+C_{8}F^{(9)}_{8}(q^2)\right],\label{WC3}
\end{eqnarray}
where the functions $h(m_{q}^{\text{pole}}, q^2)$ with $q=c, b$, and functions $F^{(7,9)}_{8}(q^2)$ are
defined in \cite{Beneke:2001at}, while the functions $F^{(7,9)}_{1,c}(q^2)$, $F^{(7,9)}_{2,c}(q^2)$ are
given in \cite{Asatryan:2001zw} for low $q^{2}$ and in \cite{Greub:2008cy} for high $q^{2}$. The quark masses appearing
in all of these functions are defined in the pole scheme.
\section{Hadronic matrix elements}\label{app:HME}
The matrix elements for the process $M_{in}\to S\ell^{+}\ell^{-}$, where the parent particle $M_{in}=B_s$ or $B$, and
the daughter particle $S$ is a scalar meson $0^{+}$, such as $S=f_{0}(980)$ or $K^{\ast}_0(1430)$, are given by
\begin{eqnarray}
\langle f_0\big(K^{\ast}_0\big)(k)|\bar s\gamma_{\mu}\gamma_{5}b|B_s\big(B\big)(p)\rangle&=&
-i\Big[f_{+}^{{f_0}({K^{\ast}_0})}(q^{2})P_{\mu}+f_{-}^{{f_0}({K^{\ast}_0})}(q^{2})q_{\mu}\Big],\label{2.2.1}\\
\langle f_0\big(K^{\ast}_0\big)(k)|\bar s i \sigma_{\mu\nu}q^{\nu}\gamma_{5}b|B_s\big(B\big)(p)\rangle&=& -i\frac{f_{T}^{{f_0}({K^{\ast}_0})}(q^{2})}{\big(m_{{B_s}(B)}+m_{{f_0}({K^{\ast}_0})}\big)}\notag
\\
&&\times\Big[q^{2}P_{\mu}-\Big(m^{2}_{{B_s}({B})}-m^{2}_{{f_0}({K^{\ast}_0})}\Big)q_{\mu}\Big],\label{2.2.2}
%\\
%\langle f_0\big(K^{\ast}_0\big)(k)|\bar s i \sigma_{\mu\nu}q^{\nu}\gamma_{5}b|\bar B_s\big(\bar B\big)(p)\rangle&=& -i\frac{f_{T}(q^{2})}{m_{B_s}\big(m_B\big)+m_{f_0}\big(m_{K^{\ast}_0}\big)}\notag
%\\
%&&\times\Big[q^{2}P_{\mu}-\Big(m^{2}_{B_s}\big(m^{2}_{B}\big)-m^{2}_{f_0}\big(m^{2}_{K^{\ast}_0}\big)\Big)\Big]q_{\mu}\label{2.2.2}
\end{eqnarray}
where $P_{\mu}= p_{\mu}+k_{\mu}$, and $q_{\mu}=p_{\mu}-k_{\mu}$. For the $B_s\to f_{0}(980)\ell^{+}\ell^{-}$ decay, $f_+^{f_0}, f_0^{f_0}, f_T^{f_0}$ form factors are used in the numerical analysis. For that $f_{-}^{f_0}(q^{2})$ form factor can be expressed as
\begin{eqnarray}
f_{-}^{f_0}(q^{2})=\frac{m^{2}_{B_s}-m^{2}_{f_0}}{q^{2}}(f_{0}^{f_0}(q^{2})-f_{+}^{f_0}(q^{2})).\label{2.2.3}
\end{eqnarray}

The matrix elements for the process $M_{in}\to P\ell^{+}\ell^{-}$, where both initial $M_{in}=B$, and
final state meson $P=K$, are pseudoscalar in nature, can be expressed as
\begin{eqnarray}
\langle K(k)|\bar s\gamma_{\mu}b|B(p)\rangle&=&f_{+}^K(q^{2})P_{\mu}+f_{-}^K(q^{2})q_{\mu},\label{2.3}\\
\langle K(k)|\bar s\sigma_{\mu\nu}q^{\nu}b|B(p)\rangle&=&-\frac{f_{T}^K(q^{2})}{m_B+m_K}
\Big[q^{2}P_{\mu}-\Big(m^{2}_{B}-m^{2}_{K}\Big)q_{\mu}\Big].\label{2.4}
\end{eqnarray}
For the $B\to K\ell^{+}\ell^{-}$ decay, $f_+^{K}, f_0^{K}, f_T^{K}$ form factors are used in the numerical analysis. Therefore,
$f_{-}^{K}(q^{2})$ form factor is decomposed using a similar expression to Eq. (\ref{2.2.3}).

For the process $M_{in}\to V\ell^{+}\ell^{-}$, where the parent particle $M_{in}=B$ or $B_s$, and
the daughter particle $V$ is a vector meson $1^{-}$, such as $V=K^{\ast}$ or $\phi$, the matrix elements
for such decays can be parameterized in terms of the form factors as
\begin{align}
\left\langle K^\ast\big(\phi\big)(k,\overline\epsilon)\left\vert \bar{s}\gamma
_{\mu }b\right\vert B\big(B_s\big)(p)\right\rangle &=\frac{2\epsilon_{\mu\nu\alpha\beta}}
{m_{B(B_s)}+m_{K^\ast(\phi)}}\overline\epsilon^{\,\ast\nu}p^{\alpha}k^{\beta}V^{K^\ast(\phi)}(q^{2}),\label{2.13a}
\\
\left\langle K^\ast\big(\phi\big)(k,\overline\epsilon)\left\vert \bar{s}\gamma_{\mu}\gamma_{5}b\right\vert
B(B_s\big)(p)\right\rangle &=i\left(m_{B(B_s)}+m_{K^\ast(\phi)}\right)g_{\mu\nu}\overline\epsilon^{\,\ast\nu}A_{1}^{K^\ast(\phi)}(q^{2})
\notag\\
&-iP_{\mu}(\overline\epsilon^{\,\ast}\cdot q)\frac{A_{2}^{K^\ast(\phi)}(q^{2})}{\left(m_{B(B_s)}+m_{K^\ast(\phi)}\right)}\notag\\
&-i\frac{2m_{K^\ast(\phi)}}{q^{2}}q_{\mu}(\overline\epsilon^{\,\ast}\cdot q)
\left[A_{3}^{K^\ast(\phi)}(q^{2})-A_{0}^{K^\ast(\phi)}(q^{2})\right],\label{2.13b}
\end{align}
where
\begin{eqnarray}
A_{3}^{K^\ast(\phi)}(q^{2})&=&\frac{m_{B(B_s)}+m_{K^\ast(\phi)}}{2m_{K^\ast(\phi)}}A_{1}^{K^\ast(\phi)}(q^{2})
-\frac{m_{B(B_s)}-m_{K^\ast(\phi)}}{2m_{K^\ast(\phi)}}A_{2}^{K^\ast(\phi)}(q^{2}),\label{A3}
\end{eqnarray}
with $A_3(0)=A_0(0)$. Here and throughout the whole study, we have used $\epsilon_{0123}=+1$ convention for the Levi-Civita tensor.
The additional form factors are the tensor form factors which can be expressed as
\begin{align}
\left\langle K^\ast\big(\phi\big)(k,\overline\epsilon)\left\vert \bar{s}i\sigma
_{\mu \nu }q^{\nu }b\right\vert B\big(B_s\big)(p)\right\rangle
&=-2\epsilon _{\mu\nu\alpha\beta}\overline\epsilon^{\,\ast\nu}p^{\alpha}k^{\beta}T_{1}^{K^\ast(\phi)}(q^{2}),\label{FF11}\\
\left\langle K^\ast\big(\phi\big)(k,\overline\epsilon )\left\vert \bar{s}i\sigma
_{\mu \nu }q^{\nu}\gamma_{5}b\right\vert B\big(B_s\big)(p)\right\rangle
&=i\Big[\left(m^2_{B(B_s)}-m^2_{K^\ast(\phi)}\right)g_{\mu\nu}\overline\epsilon^{\,\ast\nu}\notag\\
&-(\overline\epsilon^{\,\ast }\cdot q)P_{\mu}\Big]T_{2}^{K^\ast(\phi)}(q^{2})+i(\overline\epsilon^{\,\ast}\cdot q)\notag\\
&\times\left[q_{\mu}-\frac{q^{2}}{m^2_{B(B_s)}-m^2_{K^\ast(\phi)}}P_{\mu}
\right]T_{3}^{K^\ast(\phi)}(q^{2}).\label{F3}
\end{align}
%with $T_{1}(0)=T_{2}(0)$.
The relations between the form factors in \cite{Straub:2015ica}, and the form factors given in above matrix elements are
\begin{align}
A_{12}^{K^\ast(\phi)}&=\frac{\left(m_{B(B_s)}+m_{K^\ast(\phi)}\right)^2(m^2_{B(B_s)}-m^2_{K^\ast(\phi)}-q^2)
A_{1}^{K^\ast(\phi)}-\lambda A_{2}^{K^\ast(\phi)}}{16m_{B(B_s)}m^2_{K^\ast(\phi)}\left(m_{B(B_s)}+m_{K^\ast(\phi)}\right)},\notag
\\
T_{23}^{K^\ast(\phi)}&=\frac{\left(m^2_{B(B_s)}-m^2_{K^\ast(\phi)}\right)(m^2_{B(B_s)}+3m^2_{K^\ast(\phi)}-q^2)
T_{2}^{K^\ast(\phi)}-\lambda T_{3}^{K^\ast(\phi)}}{8m_{B(B_s)}m^2_{K^\ast(\phi)}\left(m_{B(B_s)}-m_{K^\ast(\phi)}\right)}.\label{relation}
\end{align}

For $M_{in}\to A\ell^{+}\ell^{-}$ decay, where $M_{in}=B$, and $M_{f}=A$, is a final state axial vector
meson $1^{+}$, such as $K_1(1270, 1400)$ meson. For this decay the matrix element can be parameterized in terms
of transition form factors as follows
\begin{eqnarray}
\langle K_1(k,\overline\epsilon)|\bar{s}\gamma_{\mu}b|B(p)\rangle&=&-\left(m_{B}+m_{K_1}\right)g_{\mu\nu}\overline\epsilon^{\,\ast\nu}V_{1}^{K_1}(q^{2})
+P_{\mu}(\overline\epsilon^{\,\ast}\cdot q)\frac{V_{2}^{K_1}(q^{2})}{\left(m_{B}+m_{K_1}\right)}\notag\\
&+&\frac{2m_{K_1}}{q^{2}}q_{\mu}(\overline\epsilon^{\,\ast}\cdot q)[V_{3}^{K_1}(q^{2})-V_{0}^{K_1}(q^{2})],\label{F1}\\
\langle K_1(k,\overline\epsilon)|\bar{s}\gamma_{\mu}\gamma_5 b|B(p)\rangle&=& \frac{2i\epsilon_{\mu\nu\alpha\beta}}{m_{B}+m_{K_1}}
\overline\epsilon^{\,\ast\nu}p^{\alpha}k^{\beta}A^{K_1}(q^{2}),\label{F2}
\end{eqnarray}
where
\begin{eqnarray}
V_{3}^{K_1}(q^{2})&=&\frac{m_{B}+m_{K_1}}{2m_{K_1}}V_{1}^{K_1}(q^{2})-\frac{m_{B}-m_{K_1}}{2m_{K_1}}V_{2}^{K_1}(q^{2}),\label{V3}\\
V_{3}^{K_1}(0)&=&V_{0}^{K_1}(0).\notag
\end{eqnarray}
The other contributions from the tensor form factors are
\begin{eqnarray}
\langle K_1(k,\overline\epsilon)|\bar si\sigma_{\mu\nu}q^{\nu}b|B(p)\rangle&=&\Big[\Big(m_{B}^{2}-m_{K_1}^{2}\Big)g_{\mu\nu}\overline\epsilon^{\,\ast\nu}
-(\overline\epsilon^{\,\ast}\cdot q)P_{\mu}\Big]T_{2}^{K_1}(q^{2})\notag\\
&+&(\overline\epsilon^{\,\ast}\cdot q)\left[q_{\mu}-\frac{q^{2}}{m^2_{B}-m^2_{K_1}}P_{\mu}\right]T_{3}^{K_1}(q^{2}),\label{T1}\\
\langle K_1(k,\overline\epsilon)|\bar si\sigma_{\mu\nu}q^{\nu}\gamma_{5}b|B(p)\rangle&=&2i\epsilon_{\mu\nu\alpha\beta}\overline\epsilon^{\,\ast\nu}p^{\alpha}k^{\beta}T_{1}^{K_1}(q^{2})\,.\label{T2}
\end{eqnarray}
%with $T_{1}^{K_1}(0)=T_{2}^{K_1}(0)$.

The matrix elements for the process $\Lambda_b\to \Lambda\ell^{+}\ell^{-}$, where both initial $M_{in}=\Lambda_b$, and
final state baryon $\Lambda$, are spin half particles, can be conveniently written in the helicity basis
\begin{align}
\left\langle \Lambda(k, s_{\Lambda})\left\vert \bar{s}\gamma_{\mu}b\right\vert \Lambda_b(p, s_{\Lambda_b})\right\rangle
&=\bar u_{\Lambda}(k, s_{\Lambda})\Big[f_t^V(q^{2})(m_{\Lambda_b}-m_{\Lambda})\frac{q_{\mu}}{q^2}\notag\\
&+f_0^V(q^2)\frac{m_{\Lambda_b}+m_{\Lambda}}{s_+}\big\{p_{\mu}+k_{\mu}-\frac{q_{\mu}}
{q^2}(m^2_{\Lambda_b}-m^2_{\Lambda})\big\}\notag
\\
&+f_{\perp}^V(q^2)\big\{\gamma_{\mu}-\frac{2m_{\Lambda}}{s_+}p_{\mu}-
\frac{2m_{\Lambda_b}}{s_+}k_{\mu}\big\}\Big]u_{\Lambda_b}(p, s_{\Lambda_b}),\label{HM1}
\\
\left\langle \Lambda(k, s_{\Lambda})\left\vert \bar{s}\gamma_{\mu}\gamma_5 b\right\vert \Lambda_b(p, s_{\Lambda_b})\right\rangle
&=-\bar u_{\Lambda}(k, s_{\Lambda})\gamma_5\Big[f_t^A(q^{2})(m_{\Lambda_b}+m_{\Lambda})\frac{q_{\mu}}{q^2}\notag\\
&+f_0^A(q^2)\frac{m_{\Lambda_b}-m_{\Lambda}}{s_-}\big\{p_{\mu}+k_{\mu}-\frac{q_{\mu}}
{q^2}(m^2_{\Lambda_b}-m^2_{\Lambda})\big\}\notag
\\
&+f_{\perp}^A(q^2)\big\{\gamma_{\mu}+\frac{2m_{\Lambda}}{s_-}p_{\mu}-
\frac{2m_{\Lambda_b}}{s_-}k_{\mu}\big\}\Big]u_{\Lambda_b}(p, s_{\Lambda_b}),\label{HM2}
\end{align}
where we have $s\pm=(m_{\Lambda_b}\pm m_\Lambda)^2-q^2$. Additionally,
\begin{align}
\langle \Lambda(k, s_{\Lambda})| \bar{s}i\sigma_{\mu\nu}\gamma^{\nu}&b | \Lambda_b(p, s_{\Lambda_b})\rangle
=-\bar u_{\Lambda}(k, s_{\Lambda})\Big[f_0^T(q^2)\frac{q^2}{s_+}\big\{p_{\mu}+k_{\mu}-\frac{q_{\mu}}
{q^2}(m^2_{\Lambda_b}-m^2_{\Lambda})\big\}\notag
\\
&+f_{\perp}^T(q^2)(m_{\Lambda_b}+m_{\Lambda})\big\{\gamma_{\mu}-\frac{2m_{\Lambda}}{s_+}p_{\mu}-
\frac{2m_{\Lambda_b}}{s_+}k_{\mu}\big\}\Big]u_{\Lambda_b}(p, s_{\Lambda_b}),\label{HM3}
\\
\langle \Lambda(k, s_{\Lambda})| \bar{s}i\sigma_{\mu\nu}\gamma^{\nu}\gamma_5 &b| \Lambda_b(p, s_{\Lambda_b})\rangle
=-\bar u_{\Lambda}(k, s_{\Lambda})\gamma_5\Big[f_0^{T_5}(q^2)\frac{q^2}{s_-}\big\{p_{\mu}+k_{\mu}-\frac{q_{\mu}}
{q^2}(m^2_{\Lambda_b}-m^2_{\Lambda})\big\}\notag
\\
&+f_{\perp}^{T_5}(q^2)(m_{\Lambda_b}-m_{\Lambda})\big\{\gamma_{\mu}+\frac{2m_{\Lambda}}{s_-}p_{\mu}-
\frac{2m_{\Lambda_b}}{s_-}k_{\mu}\big\}\Big]u_{\Lambda_b}(p, s_{\Lambda_b}).\label{HM4}
\end{align}
\section{Details on the kinematics}
\subsection{Kinematics}
\begin{figure}[t]
\begin{center}
\includegraphics[scale=0.40]{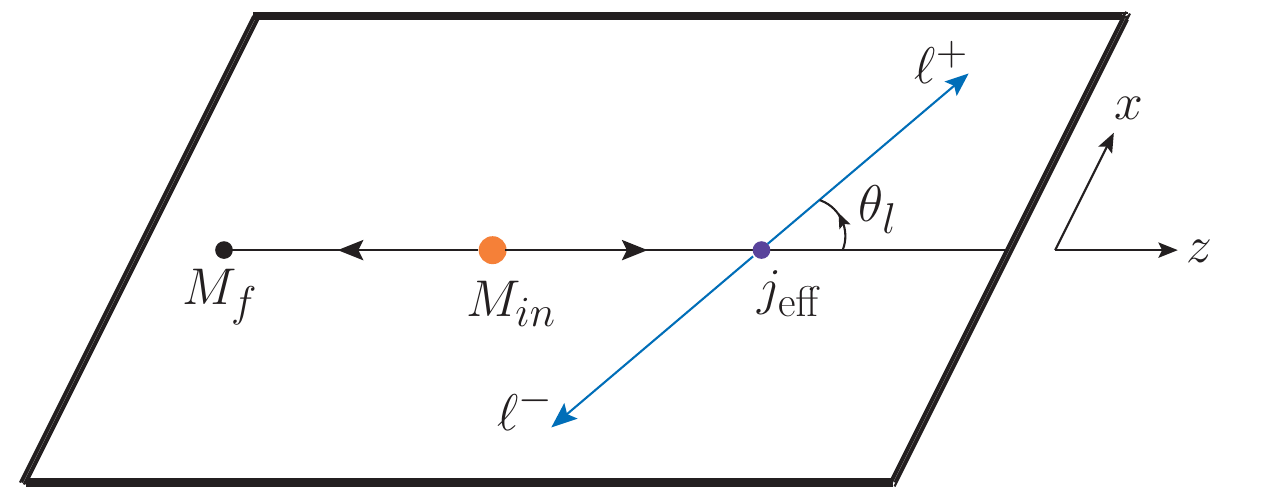}
\end{center}
\caption{Kinematics of the $M_{in}\to M_f\ell^{+}\ell^{-}$ decay.}
\label{figlab}
\end{figure}

The decay $M_{in}\to M_{f} \ell^+\ell^-$ can be conveniently regarded as a quasi-two-body decay
with $M_{in}\to M_{f}j_{\text{eff}}$ followed by $j_{\text{eff}}\to \ell^+\ell^-$, where effective current $j_{\text{eff}}$, represents
the off-shell boson. The polarization vectors of $j_{\text{eff}}$ satisfy
the orthonormality and completeness relations as discussed in section \ref{secHelicity}. With $M_{in}(p)\to M_{f}(k)\left(j_{\text{eff}}(q)\to \ell^+(p_1)\ell^-(p_2)\right)$, we define momenta in the rest frame of the parent particle $M_{in}$ as
\begin{eqnarray}
p^{\mu}=(m_{in}, 0, 0, 0), \qquad k^{\mu}=(E_{f}, 0, 0, -|\vec{k}|), \qquad q^{\mu}=(q^0, 0, 0, +|\vec{k}|),\label{A01}
\end{eqnarray}
where we choose daughter particle $M_f$ to be moving along the negative $z$ direction, and
\begin{eqnarray}
q^0=\frac{m^2_{in}-m^2_{f}+q^2}{2 m_{in}}, \qquad E_f=\frac{m^2_{in}+m^2_{f}-q^2}{2 m_{in}},
\qquad |\vec{k}|=\frac{\sqrt{\lambda(m^2_{in}, m^2_{f}, q^2)}}{2 m_{in}},\label{A02}
\end{eqnarray}
where $\lambda(m^2_{in}, m^2_{f}, q^2)$ is the K{\"a}ll{\'e}n function
\begin{eqnarray}
\lambda(a, b, c)=a^2+b^2+c^2-2(ab+ac+bc).\label{A03}
\end{eqnarray}
In the dilepton rest frame, considering $j_{\text{eff}}$ decaying in the $x-z$ plane, and $\ell^+(p_1)$ lepton making angle
$\theta_l$ with the $z-$axis (see figure \ref{figlab}),
\begin{eqnarray}
p^{\mu}_1&=& (E_l, |\vec{p_l}|\sin\theta_l, 0, |\vec{p_l}|\cos\theta_l),\notag
\\
p^{\mu}_2&=& (E_l, -|\vec{p_l}|\sin\theta_l, 0, -|\vec{p_l}|\cos\theta_l),\label{A04}
\end{eqnarray}
with
\begin{eqnarray}
E_l=\frac{\sqrt{q^2}}{2}, \qquad |\vec{p_l}|=\frac{\sqrt{q^2}}{2}\beta_l, \qquad  \beta_l=\sqrt{1-\frac{4m_l^2}{q^2}}.\label{A05}
\end{eqnarray}
\subsection{Polarization conventions}
In the $M_{in}$ rest frame, the polarization four-vectors of the effective current $(j_{\text{eff}})$, that decays to
dilepton pair are
\begin{eqnarray}
\varepsilon^{\mu}(t)=\frac{1}{\sqrt{q^2}}(q^0, 0, 0, |\vec{k}|),\;\;
\varepsilon^{\mu}(\pm)=\frac{1}{\sqrt{2}}(0, \mp1, -i, 0),\quad \varepsilon^{\mu}(0)=\frac{1}{\sqrt{q^2}}(|\vec{k}|, 0, 0, q^0),\label{B01}
\end{eqnarray}
and in the dilepton pair rest frame the transverse polarizations of $j_{\text{eff}}$ remain same, while the time like and longitudinal polarizations
read
\begin{eqnarray}
\varepsilon^{\mu}(t)=(1, 0, 0, 0),\qquad  \varepsilon^{\mu}(0)=(0, 0, 0, 1).\label{B02}
\end{eqnarray}
Similarly, when the final state is vector or axial-vector particle, the polarization four-vectors of $V(A)$ state moving
along the negative $z$ direction, in the $M_{in}$ rest frame are
\begin{eqnarray}
\overline\epsilon^{\mu}(\pm)=\frac{1}{\sqrt{2}}(0, \pm1, -i, 0),\quad \overline\epsilon^{\mu}(0)=\frac{1}{m_f}(|\vec{k}|, 0, 0, E_f).\label{B03}
\end{eqnarray}
Transverse polarizations of $V(A)$ in its own rest frame remain same, whereas the longitudinal polarization reads
\begin{eqnarray}
\overline\epsilon^{\mu}(0)=(0, 0, 0, -1).\label{B04}
\end{eqnarray}
\section{Numerical inputs}
In Table~\ref{tab:Numinputs} we give the numerical values of the input parameters used in our study.
\begin{table*}[!htbp]
\renewcommand{\arraystretch}{1}
	\begin{center}
			\begin{tabular}{|l@{\hspace{0.3cm}} r|l@{\hspace{0.3cm}} r|}
				\hline
				%\multicolumn{2}{|c|}{Input parameters}  & \multicolumn{2}{c|}{Input parameters}  \\
				%\hline
                $G_{F}=1.1663787\times 10^{-5}$ GeV$^{-2}$ & \cite{Zyla:2020zbs}                          &$m_{B}=5.279$ GeV       & \cite{Zyla:2020zbs}\\
                $|V_{tb}V_{ts}^{\ast}|=0.0397^{+0.0008}_{-0.0006}$ & \cite{Zyla:2020zbs}                 & $m_{B_s}=5.367$ GeV     & \cite{Zyla:2020zbs} \\
                $m_b=4.18^{+0.03}_{-0.02}$ GeV             & \cite{Zyla:2020zbs}        &  $\tau_{B}=(1.519\pm0.004)\times 10^{-12}$ s  &\cite{Zyla:2020zbs} \\
                $\alpha(m_b)=1/133.28$                 & \cite{Detmold:2016pkz}        & $\tau_{B_s}=(1.515\pm0.004)\times 10^{-12}$ s  & \cite{Zyla:2020zbs}\\
                $\alpha_s(m_b)= 0.2233$                & \cite{Detmold:2016pkz}                           &$m_{f_0}=0.990$ GeV        & \cite{Zyla:2020zbs} \\
                $m_e=0.0005$ GeV                             & \cite{Zyla:2020zbs}                &$m_{K^{\ast}_0}=1.425$ GeV       & \cite{Zyla:2020zbs} \\
                $m_{\mu}=0.106$ GeV                          & \cite{Zyla:2020zbs}                        &$m_{K}=0.498$ GeV       & \cite{Zyla:2020zbs}  \\
                 $m_b^{\text{pole}}=4.91\pm0.12$ GeV      & \cite{Ali:2013zfa}                   &$m_{K^{\ast}}=0.896$ GeV       & \cite{Zyla:2020zbs} \\
                $m_c^{\text{pole}}=1.77\pm0.14$ GeV      & \cite{Ali:2013zfa}                   &$m_{\phi}=1.020$ GeV        & \cite{Zyla:2020zbs}\\
                $\mu_b=5$ GeV                            & \cite{Du:2015tda}           &$m_{K_{1A}}=1.31$ GeV         & \cite{Yang:2007zt} \\
                $m_{\Lambda_b}=5.619$ GeV                & \cite{Zyla:2020zbs}         &$m_{K_{1B}}=1.34$ GeV         & \cite{Yang:2007zt} \\
                $m_{\Lambda}=1.116$ GeV    & \cite{Zyla:2020zbs} &$m_{K_1(1270)}=1.272$ GeV        & \cite{Tanabashi:2018oca} \\
                 $\tau_{\Lambda_b}=(1.471\pm0.009)\times 10^{-12}$ s  & \cite{Zyla:2020zbs} &$m_{K_1(1400)}=1.403$ GeV      & \cite{Zyla:2020zbs}  \\
                \hline
	\end{tabular}		
    \caption{Default values of the used input parameters. Values of some parameters are strongly scale
dependent, but most of these parameters cancel in the LFUV ratios.}\label{tab:Numinputs}
	\end{center}
\end{table*}
\section{$\Lambda_b \to \Lambda$ spinor bilinears}\label{spinorele}
To calculate the hadronic helicity amplitudes for $\Lambda_b \to \Lambda \ell^+\ell^-$ decay, we use the spinor
representations given in \cite{Boer:2014kda,Haber:1994pe}. For scalar and pseudo-scalar currents, we get
\begin{eqnarray}
&&\bar u_{\Lambda}(k, \pm1/2)u_{\Lambda_b}(p, \pm1/2)=0,\notag
\\
&&\bar u_{\Lambda}(k, \pm1/2)u_{\Lambda_b}(p, \mp1/2)=\pm\sqrt{s_+},\notag
\\
&&\bar u_{\Lambda}(k, \pm1/2)\gamma_5 u_{\Lambda_b}(p, \pm1/2)=0\notag
\\
&&\bar u_{\Lambda}(k, \pm1/2)\gamma_5 u_{\Lambda_b}(p, \pm1/2)=-\sqrt{s_-},\label{C01}
\end{eqnarray}
and for vector and axial-vector currents, we obtain
\begin{eqnarray}
&&\bar u_{\Lambda}(k, \pm1/2)\gamma^{\mu}u_{\Lambda_b}(p, \pm1/2)=\mp\sqrt{2s_-}\epsilon^{\mu}(\pm),\notag
\\
&&\bar u_{\Lambda}(k, \pm1/2)\gamma^{\mu}u_{\Lambda_b}(p, \mp1/2)=\pm(\sqrt{s_+}, 0, 0, -\sqrt{s_-}),\notag
\\
&&\bar u_{\Lambda}(k, \pm1/2)\gamma^{\mu}\gamma_5 u_{\Lambda_b}(p, \pm1/2)=-\sqrt{2s_+}\epsilon^{\mu}(\pm)\notag
\\
&&\bar u_{\Lambda}(k, \pm1/2)\gamma^{\mu}\gamma_5 u_{\Lambda_b}(p, \mp1/2)=(\sqrt{s_-}, 0, 0, -\sqrt{s_+}).\label{C02}
\end{eqnarray}
\section{Predicted values of the LFUV ratios}
\label{app:LFUV}
In this appendix, we give the predicted central values with errors for the various LFUV ratios.
\begin{table*}[!htbp]
	\begin{center}
    \setlength\tabcolsep{3.2pt}
			\begin{tabular}{|ccccc|}
				\hline
				Observable & Scenario & $q^2$/GeV$^2: [0.045,1]$ & $q^2$/GeV$^2: [1,6]$  & $q^2$/GeV$^2: [14,q^2_{\text{max}}]$ \\
				\hline
				$R_{f_0}$& SM & $0.977_{-0.036}^{+0.051}$ & $1.000_{-0.005}^{+0.007}$ & $1.006_{-0.005}^{+0.006}$ \\
                \hline	
                $R_{f_0}$& S1  & $0.766_{-0.037(-0.023)}^{+0.052(+0.025)}$ & $0.785_{-0.006(-0.024)}^{+0.009(+0.026)}$ & $0.789_{-0.006(-0.024)}^{+0.008(+0.026)}$ \\
                $R_{f_0}$& S2 & $0.774_{-0.028(-0.025)}^{+0.039(+0.022)}$ & $0.792_{-0.004(-0.025)}^{+0.005(+0.022)}$ & $0.802_{-0.005(-0.025)}^{+0.004(+0.022)}$ \\
                $R_{f_0}$& S3 & $0.783_{-0.029(-0.078)}^{+0.040(+0.078)}$ & $0.801_{-0.004(-0.080)}^{+0.005(+0.080)}$ & $0.811_{-0.005(-0.078)}^{+0.004(+0.078)}$ \\
                $R_{f_0}$& S4 & $0.802_{-0.029(-0.093)}^{+0.039(+0.093)}$ & $0.821_{-0.005(-0.096)}^{+0.006(+0.096)}$ & $0.835_{-0.009(-0.097)}^{+0.007(+0.097)}$ \\
                $R_{f_0}$& S5 & $0.824_{-0.038(-0.032)}^{+0.053(+0.032)}$ & $0.843_{-0.005(-0.033)}^{+0.007(+0.033)}$ & $0.852_{-0.004(-0.032)}^{+0.005(+0.032)}$ \\
                $R_{f_0}$& S6 & $0.738_{-0.035(-0.030)}^{+0.049(+0.032)}$ & $0.755_{-0.006(-0.031)}^{+0.008(+0.032)}$ & $0.760_{-0.006(-0.031)}^{+0.007(+0.033)}$ \\
                \hline
	\end{tabular}		
	\caption{SM and NP predictions for the LFUV ratio $R_{f_0}$ in different bins. The first errors listed are due to the uncertainties of the form factors, and the second errors are due to the $1\sigma$ range of the best-fit Wilson coefficients in different NP scenarios.}\label{tab:Rf0bin}
	\end{center}
\end{table*}
\begin{table*}[!htbp]
	\begin{center}
    \setlength\tabcolsep{1.3pt}
			\begin{tabular}{|ccccc|}
				\hline
				Observable & Scenario & $q^2$/GeV$^2: [0.045,1]$ & $q^2$/GeV$^2: [1,6]$  & $q^2$/GeV$^2: [14,q^2_{\text{max}}]$ \\
				\hline
				$R_{K^*_0}$& SM & $0.977\pm0.001$ & $1.000\pm0.001$ & $1.012\pm0.014$ \\
                \hline	
                $R_{K^*_0}$& S1  & $0.759\pm0.001_{(-0.025)}^{(+0.027)}$ & $0.778\pm0.001_{(-0.026)}^{(+0.028)}$ & $0.791\pm0.014_{(-0.026)}^{(+0.028)}$ \\
                $R_{K^*_0}$& S2 & $0.795\pm0.003_{(-0.022)}^{(+0.019)}$ & $0.814\pm0.003_{(-0.023)}^{(+0.020)}$ & $0.829\pm0.011_{(-0.022)}^{(+0.020)}$ \\
                $R_{K^*_0}$& S3 & $0.805\pm0.003_{(-0.070)}^{(+0.070)}$ & $0.824\pm0.003_{(-0.072)}^{(+0.072)}$ & $0.838\pm0.011_{(-0.071)}^{(+0.071)}$ \\
                $R_{K^*_0}$& S4 & $0.839\pm0.004_{(-0.098)}^{(+0.098)}$ & $0.859\pm0.004_{(-0.100)}^{(+0.100)}$ & $0.876\pm0.011_{(-0.101)}^{(+0.101)}$ \\
                $R_{K^*_0}$& S5 & $0.837\pm0.002_{(-0.029)}^{(+0.029)}$ & $0.857\pm0.002_{(-0.029)}^{(+0.029)}$ & $0.873\pm0.015_{(-0.029)}^{(+0.029)}$ \\
                $R_{K^*_0}$& S6 & $0.733\pm0.001_{(-0.031)}^{(+0.033)}$ & $0.751\pm0.001_{(-0.032)}^{(+0.034)}$ & $0.764\pm0.013_{(-0.032)}^{(+0.034)}$ \\
                \hline
	\end{tabular}		
	\caption{SM and NP predictions for the LFUV ratio $R_{K^*_0}$ in different bins. The first errors listed are due to the uncertainties of the form factors, and the second errors are due to the $1\sigma$ range of the best-fit Wilson coefficients in different NP scenarios.}\label{tab:RKst0bin}
	\end{center}
\end{table*}
\begin{table*}[!htbp]
	\begin{center}
    \setlength\tabcolsep{4.8pt}
			\begin{tabular}{|ccccc|}
				\hline
				 Scenario & Observable & $q^2$/GeV$^2: [14,q^2_{\text{max}}]$ & Observable  & $q^2$/GeV$^2: [14,q^2_{\text{max}}]$ \\
				\hline
				SM & $R_{K}$ & $1.002_{-0.002}^{+0.002}$ & $R_{K^*}$ & $0.998\pm0.000$ \\
                \hline	
                S1  & $R_{K}$ & $0.785_{-0.003(-0.024)}^{+0.004(+0.026)}$ & $R_{K^*}$ & $0.785\pm0.002_{(-0.023)}^{(+0.025)}$ \\
                S2 & $R_{K}$ & $0.800_{-0.005(-0.024)}^{+0.003(+0.021)}$ & $R_{K^*}$ & $0.790\pm0.003_{(-0.025)}^{(+0.022)}$ \\
                S3 & $R_{K}$ & $0.810_{-0.005(-0.077)}^{+0.003(+0.077)}$ & $R_{K^*}$ & $0.800\pm0.003_{(-0.080)}^{(+0.080)}$ \\
                S4 & $R_{K}$ & $0.834_{-0.009(-0.097)}^{+0.006(+0.097)}$ & $R_{K^*}$ & $0.818\pm0.005_{(-0.094)}^{(+0.094)}$ \\
                S5 & $R_{K}$ & $0.849_{-0.003(-0.032)}^{+0.002(+0.032)}$ & $R_{K^*}$ & $0.841\pm0.001_{(-0.033)}^{(+0.033)}$ \\
                S6 & $R_{K}$ & $0.791_{-0.003(-0.026)}^{+0.004(+0.028)}$ & $R_{K^*}$ & $0.762\pm0.003_{(-0.028)}^{(+0.030)}$ \\
                \hline
                SM & $R_{{K^*}^L}$ & $0.999\pm0.000$ & $R_{{K^*}^T}$ & $0.998\pm0.000$ \\
                \hline	
                S1 & $R_{{K^*}^L}$  & $0.783\pm0.002_{(-0.024)}^{(+0.026)}$ & $R_{{K^*}^T}$ & $0.786\pm0.003_{(-0.023)}^{(+0.025)}$ \\
                S2 & $R_{{K^*}^L}$ & $0.794\pm0.003_{(-0.025)}^{(+0.022)}$ & $R_{{K^*}^T}$ & $0.788\pm0.004_{(-0.025)}^{(+0.022)}$ \\
                S3 & $R_{{K^*}^L}$ & $0.803\pm0.003_{(-0.079)}^{(+0.079)}$ & $R_{{K^*}^T}$ & $0.798\pm0.004_{(-0.080)}^{(+0.080)}$ \\
                S4 & $R_{{K^*}^L}$ & $0.824\pm0.006_{(-0.095)}^{(+0.095)}$ & $R_{{K^*}^T}$ & $0.814\pm0.008_{(-0.094)}^{(+0.094)}$ \\
                S5 & $R_{{K^*}^L}$ & $0.844\pm0.002_{(-0.032)}^{(+0.032)}$ & $R_{{K^*}^T}$ & $0.840\pm0.002_{(-0.033)}^{(+0.033)}$ \\
                S6 & $R_{{K^*}^L}$ & $0.755\pm0.002_{(-0.030)}^{(+0.032)}$ & $R_{{K^*}^T}$ & $0.766\pm0.005_{(-0.028)}^{(+0.029)}$ \\
                \hline
	\end{tabular}		
	\caption{SM and NP predictions for the LFUV ratios $R_{K}$, $R_{{K^*}^{(L,\,T)}}$ in the high $q^2$ bin. The first errors listed are due to the uncertainties of the form factors, and the second errors are due to the $1\sigma$ range of the best-fit Wilson coefficients in different NP scenarios.}\label{tab:RKKstbin}
	\end{center}
\end{table*}
\begin{table*}[!htbp]
	\begin{center}
    \setlength\tabcolsep{1.3pt}
			\begin{tabular}{|ccccc|}
				\hline
				Observable & Scenario & $q^2$/GeV$^2: [0.045,1]$ & $q^2$/GeV$^2: [1,6]$  & $q^2$/GeV$^2: [14,q^2_{\text{max}}]$ \\
				\hline
				$R_{\phi}$& SM & $0.927\pm0.004$ & $0.997\pm0.002$ & $0.998\pm0.000$ \\
                \hline	
                $R_{\phi}$& S1  & $0.872\pm0.013_{(-0.005)}^{(+0.006)}$ & $0.826\pm0.008_{(-0.018)}^{(+0.019)}$ & $0.784\pm0.001_{(-0.024)}^{(+0.025)}$ \\
                $R_{\phi}$& S2 & $0.858\pm0.010_{(-0.008)}^{(+0.007)}$ & $0.800\pm0.002_{(-0.024)}^{(+0.021)}$ & $0.791\pm0.002_{(-0.025)}^{(+0.022)}$ \\
                $R_{\phi}$& S3 & $0.866\pm0.009_{(-0.026)}^{(+0.026)}$ & $0.812\pm0.002_{(-0.076)}^{(+0.076)}$ & $0.801\pm0.002_{(-0.079)}^{(+0.079)}$ \\
                $R_{\phi}$& S4 & $0.863\pm0.007_{(-0.024)}^{(+0.024)}$ & $0.811\pm0.003_{(-0.076)}^{(+0.076)}$ & $0.819\pm0.003_{(-0.094)}^{(+0.094)}$ \\
                $R_{\phi}$& S5 & $0.881\pm0.008_{(-0.010)}^{(+0.010)}$ & $0.856\pm0.003_{(-0.030)}^{(+0.030)}$ & $0.842\pm0.001_{(-0.032)}^{(+0.032)}$ \\
                $R_{\phi}$& S6 & $0.867\pm0.013_{(-0.006)}^{(+0.007)}$ & $0.806\pm0.009_{(-0.022)}^{(+0.023)}$ & $0.760\pm0.002_{(-0.029)}^{(+0.031)}$ \\
                \hline
                $R_{\phi^L}$& SM & $0.974\pm0.016$ & $1.000\pm0.002$ & $0.999\pm0.000$ \\
                \hline	
                $R_{\phi^L}$& S1  & $0.763\pm0.016_{(-0.023)}^{(+0.025)}$ & $0.783\pm0.003_{(-0.024)}^{(+0.026)}$ & $0.783\pm0.002_{(-0.024)}^{(+0.026)}$ \\
                $R_{\phi^L}$& S2 & $0.772\pm0.012_{(-0.024)}^{(+0.021)}$ & $0.792\pm0.001_{(-0.025)}^{(+0.022)}$ & $0.795\pm0.003_{(-0.025)}^{(+0.022)}$ \\
                $R_{\phi^L}$& S3 & $0.781\pm0.012_{(-0.077)}^{(+0.078)}$ & $0.801\pm0.001_{(-0.079)}^{(+0.079)}$ & $0.804\pm0.003_{(-0.078)}^{(+0.078)}$ \\
                $R_{\phi^L}$& S4 & $0.801\pm0.012_{(-0.094)}^{(+0.094)}$ & $0.822\pm0.002_{(-0.096)}^{(+0.096)}$ & $0.826\pm0.005_{(-0.096)}^{(+0.096)}$ \\
                $R_{\phi^L}$& S5 & $0.821\pm0.016_{(-0.032)}^{(+0.032)}$ & $0.843\pm0.002_{(-0.033)}^{(+0.033)}$ & $0.844\pm0.001_{(-0.032)}^{(+0.032)}$ \\
                $R_{\phi^L}$& S6 & $0.734\pm0.015_{(-0.029)}^{(+0.032)}$ & $0.754\pm0.003_{(-0.031)}^{(+0.033)}$ & $0.754\pm0.002_{(-0.031)}^{(+0.032)}$ \\
                \hline
                $R_{\phi^T}$& SM & $0.897\pm0.000$ & $0.985\pm0.000$ & $0.998\pm0.000$ \\
                \hline	
                $R_{\phi^T}$& S1  & $0.940\pm0.001_{(-0.006)}^{(+0.006)}$ & $1.020\pm0.014_{(-0.011)}^{(+0.012)}$ & $0.785\pm0.001_{(-0.023)}^{(+0.025)}$ \\
                $R_{\phi^T}$& S2 & $0.912\pm0.000_{(-0.003)}^{(+0.003)}$ & $0.834\pm0.008_{(-0.026)}^{(+0.023)}$ & $0.789\pm0.002_{(-0.025)}^{(+0.022)}$ \\
                $R_{\phi^T}$& S3 & $0.911\pm0.000_{(-0.010)}^{(+0.010)}$ & $0.856\pm0.008_{(-0.085)}^{(+0.085)}$ & $0.798\pm0.002_{(-0.080)}^{(+0.080)}$ \\
                $R_{\phi^T}$& S4 & $0.895\pm0.001_{(-0.017)}^{(+0.017)}$ & $0.769\pm0.004_{(-0.046)}^{(+0.046)}$ & $0.816\pm0.003_{(-0.094)}^{(+0.094)}$ \\
                $R_{\phi^T}$& S5 & $0.910\pm0.000_{(-0.004)}^{(+0.004)}$ & $0.905\pm0.006_{(-0.032)}^{(+0.032)}$ & $0.840\pm0.001_{(-0.033)}^{(+0.033)}$ \\
                $R_{\phi^T}$& S6 & $0.943\pm0.001_{(-0.007)}^{(+0.007)}$ & $1.025\pm0.015_{(-0.014)}^{(+0.015)}$ & $0.764\pm0.002_{(-0.028)}^{(+0.030)}$ \\
                \hline
	\end{tabular}		
	\caption{SM and NP predictions for the LFUV ratios $R_{\phi^{(L,\,T)}}$ in different bins. The first errors listed are due to the uncertainties of the form factors, and the second errors are due to the $1\sigma$ range of the best-fit Wilson coefficients in different NP scenarios.}\label{tab:Rphibin}
	\end{center}
\end{table*}
\begin{table*}[!htbp]
	\begin{center}
    \setlength\tabcolsep{3.2pt}
			\begin{tabular}{|ccccc|}
				\hline
				Observable & Scenario & $q^2$/GeV$^2: [0.045,1]$ & $q^2$/GeV$^2: [1,6]$  & $q^2$/GeV$^2: [14,q^2_{\text{max}}]$ \\
				\hline
				$R_{K_1(1270)}$& SM & $0.922_{-0.016}^{+0.018}$ & $0.995_{-0.005}^{+0.008}$ & $0.998_{-0.000}^{+0.000}$ \\
                \hline	
                $R_{K_1(1270)}$& S1  & $0.863_{-0.051(-0.006)}^{+0.069(+0.006)}$ & $0.819_{-0.024(-0.018)}^{+0.048(+0.020)}$ & $0.782_{-0.002(-0.024)}^{+0.003(+0.026)}$ \\
                $R_{K_1(1270)}$& S2 & $0.853_{-0.038(-0.009)}^{+0.053(+0.007)}$ & $0.794_{-0.005(-0.025)}^{+0.011(+0.022)}$ & $0.793_{-0.005(-0.025)}^{+0.003(+0.022)}$ \\
                $R_{K_1(1270)}$& S3 & $0.860_{-0.035(-0.026)}^{+0.047(+0.026)}$ & $0.805_{-0.006(-0.078)}^{+0.013(+0.078)}$ & $0.802_{-0.005(-0.078)}^{+0.003(+0.078)}$ \\
                $R_{K_1(1270)}$& S4 & $0.859_{-0.029(-0.026)}^{+0.039(+0.026)}$ & $0.805_{-0.011(-0.078)}^{+0.009(+0.078)}$ & $0.823_{-0.009(-0.095)}^{+0.006(+0.095)}$ \\
                $R_{K_1(1270)}$& S5 & $0.875_{-0.034(-0.010)}^{+0.044(+0.010)}$ & $0.850_{-0.009(-0.031)}^{+0.019(+0.031)}$ & $0.842_{-0.003(-0.032)}^{+0.002(+0.032)}$ \\
                $R_{K_1(1270)}$& S6 & $0.864_{-0.054(-0.007)}^{+0.073(+0.007)}$ & $0.823_{-0.023(-0.020)}^{+0.047(+0.021)}$ & $0.788_{-0.002(-0.026)}^{+0.004(+0.027)}$ \\
                \hline
                $R_{K_1^L(1270)}$& SM & $0.963_{-0.049}^{+0.103}$ & $0.999_{-0.007}^{+0.014}$ & $0.998_{-0.000}^{+0.000}$ \\
                \hline	
                $R_{K_1^L(1270)}$& S1  & $0.749_{-0.050(-0.024)}^{+0.105(+0.026)}$ & $0.781_{-0.008(-0.026)}^{+0.018(+0.028)}$ & $0.782_{-0.002(-0.024)}^{+0.003(+0.026)}$ \\
                $R_{K_1^L(1270)}$& S2 & $0.773_{-0.039(-0.023)}^{+0.081(+0.020)}$ & $0.793_{-0.006(-0.025)}^{+0.007(+0.022)}$ & $0.794_{-0.005(-0.025)}^{+0.003(+0.022)}$ \\
                $R_{K_1^L(1270)}$& S3 & $0.781_{-0.039(-0.073)}^{+0.082(+0.073)}$ & $0.802_{-0.006(-0.079)}^{+0.007(+0.079)}$ & $0.804_{-0.005(-0.078)}^{+0.003(+0.078)}$ \\
                $R_{K_1^L(1270)}$& S4 & $0.809_{-0.040(-0.095)}^{+0.084(+0.095)}$ & $0.824_{-0.009(-0.096)}^{+0.007(+0.096)}$ & $0.826_{-0.009(-0.096)}^{+0.006(+0.096)}$ \\
                $R_{K_1^L(1270)}$& S5 & $0.816_{-0.053(-0.030)}^{+0.110(+0.030)}$ & $0.842_{-0.007(-0.032)}^{+0.013(+0.032)}$ & $0.843_{-0.003(-0.032)}^{+0.002(+0.032)}$ \\
                $R_{K_1^L(1270)}$& S6 & $0.754_{-0.054(-0.026)}^{+0.113(+0.028)}$ & $0.788_{-0.010(-0.026)}^{+0.020(+0.028)}$ & $0.789_{-0.002(-0.026)}^{+0.004(+0.028)}$ \\
                \hline
                $R_{K_1^T(1270)}$& SM & $0.893_{-0.003}^{+0.004}$ & $0.984_{-0.000}^{+0.001}$ & $0.998_{-0.000}^{+0.000}$ \\
                \hline	
                $R_{K_1^T(1270)}$& S1  & $0.947_{-0.007(-0.007)}^{+0.005(+0.008)}$ & $0.939_{-0.046(-0.001)}^{+0.067(+0.001)}$ & $0.783_{-0.002(-0.024)}^{+0.003(+0.025)}$ \\
                $R_{K_1^T(1270)}$& S2 & $0.911_{-0.001(-0.004)}^{+0.001(+0.004)}$ & $0.796_{-0.017(-0.027)}^{+0.030(+0.024)}$ & $0.792_{-0.005(-0.025)}^{+0.003(+0.022)}$ \\
                $R_{K_1^T(1270)}$& S3 & $0.910_{-0.001(-0.012)}^{+0.001(+0.012)}$ & $0.814_{-0.020(-0.087)}^{+0.033(+0.087)}$ & $0.801_{-0.005(-0.079)}^{+0.003(+0.079)}$ \\
                $R_{K_1^T(1270)}$& S4 & $0.889_{-0.004(-0.021)}^{+0.005(+0.021)}$ & $0.752_{-0.007(-0.048)}^{+0.015(+0.048)}$ & $0.821_{-0.009(-0.095)}^{+0.006(+0.095)}$ \\
                $R_{K_1^T(1270)}$& S5 & $0.909_{-0.001(-0.005)}^{+0.001(+0.005)}$ & $0.872_{-0.018(-0.033)}^{+0.028(+0.033)}$ & $0.842_{-0.003(-0.032)}^{+0.002(+0.032)}$ \\
                $R_{K_1^T(1270)}$& S6 & $0.950_{-0.007(-0.009)}^{+0.005(+0.008)}$ & $0.941_{-0.047(-0.002)}^{+0.068(+0.002)}$ & $0.787_{-0.002(-0.026)}^{+0.004(+0.027)}$ \\
                \hline
	\end{tabular}		
	\caption{SM and NP predictions for the LFUV ratios $R_{K_1^{(L,\,T)}(1270)}$, with $\theta_{K_1}=-34^{\circ}$, in different bins. The first errors listed are due to the uncertainties of the form factors, and the second errors are due to the $1\sigma$ range of the best-fit Wilson coefficients in different NP scenarios.}\label{tab:RK11270bin}
	\end{center}
\end{table*}
\begin{table*}[!htbp]
	\begin{center}
    \setlength\tabcolsep{3.2pt}
			\begin{tabular}{|ccccc|}
				\hline
				Observable & Scenario & $q^2$/GeV$^2: [0.045,1]$ & $q^2$/GeV$^2: [1,6]$  & $q^2$/GeV$^2: [14,q^2_{\text{max}}]$ \\
				\hline
				$R_{K_1(1400)}$& SM & $0.913_{-0.012}^{+0.011}$ & $0.994_{-0.004}^{+0.006}$ & $0.998_{-0.000}^{+0.000}$ \\
                \hline	
                $R_{K_1(1400)}$& S1  & $0.904_{-0.054(-0.000)}^{+0.057(+0.000)}$ & $0.838_{-0.034(-0.015)}^{+0.062(+0.016)}$ & $0.782_{-0.002(-0.024)}^{+0.003(+0.026)}$ \\
                $R_{K_1(1400)}$& S2 & $0.873_{-0.042(-0.006)}^{+0.046(+0.005)}$ & $0.792_{-0.006(-0.025)}^{+0.014(+0.022)}$ & $0.793_{-0.005(-0.025)}^{+0.003(+0.022)}$ \\
                $R_{K_1(1400)}$& S3 & $0.879_{-0.037(-0.018)}^{+0.040(+0.018)}$ & $0.804_{-0.008(-0.079)}^{+0.017(+0.079)}$ & $0.802_{-0.005(-0.078)}^{+0.003(+0.078)}$ \\
                $R_{K_1(1400)}$& S4 & $0.864_{-0.031(-0.010)}^{+0.034(+0.010)}$ & $0.793_{-0.014(-0.071)}^{+0.013(+0.071)}$ & $0.823_{-0.009(-0.095)}^{+0.006(+0.095)}$ \\
                $R_{K_1(1400)}$& S5 & $0.889_{-0.033(-0.007)}^{+0.035(+0.007)}$ & $0.853_{-0.011(-0.031)}^{+0.022(+0.031)}$ & $0.842_{-0.003(-0.032)}^{+0.002(+0.032)}$ \\
                $R_{K_1(1400)}$& S6 & $0.906_{-0.057(-0.001)}^{+0.060(+0.001)}$ & $0.842_{-0.034(-0.016)}^{+0.061(+0.017)}$ & $0.788_{-0.002(-0.026)}^{+0.004(+0.028)}$ \\
                \hline
                $R_{K_1^L(1400)}$& SM & $0.973_{-0.067}^{+0.159}$ & $0.998_{-0.007}^{+0.015}$ & $0.998_{-0.000}^{+0.000}$ \\
                \hline	
                $R_{K_1^L(1400)}$& S1  & $0.778_{-0.079(-0.021)}^{+0.192(+0.022)}$ & $0.783_{-0.011(-0.024)}^{+0.022(+0.025)}$ & $0.782_{-0.002(-0.024)}^{+0.003(+0.026)}$ \\
                $R_{K_1^L(1400)}$& S2 & $0.761_{-0.049(-0.026)}^{+0.122(+0.023)}$ & $0.789_{-0.007(-0.025)}^{+0.007(+0.022)}$ & $0.794_{-0.005(-0.025)}^{+0.003(+0.022)}$ \\
                $R_{K_1^L(1400)}$& S3 & $0.770_{-0.050(-0.082)}^{+0.125(+0.082)}$ & $0.798_{-0.007(-0.080)}^{+0.007(+0.080)}$ & $0.803_{-0.005(-0.078)}^{+0.003(+0.078)}$ \\
                $R_{K_1^L(1400)}$& S4 & $0.775_{-0.046(-0.085)}^{+0.104(+0.085)}$ & $0.816_{-0.012(-0.095)}^{+0.010(+0.095)}$ & $0.824_{-0.009(-0.096)}^{+0.006(+0.096)}$ \\
                $R_{K_1^L(1400)}$& S5 & $0.813_{-0.068(-0.033)}^{+0.163(+0.033)}$ & $0.840_{-0.007(-0.033)}^{+0.013(+0.033)}$ & $0.843_{-0.003(-0.032)}^{+0.002(+0.032)}$ \\
                $R_{K_1^L(1400)}$& S6 & $0.786_{-0.083(-0.022)}^{+0.200(+0.024)}$ & $0.791_{-0.012(-0.026)}^{+0.025(+0.027)}$ & $0.789_{-0.002(-0.026)}^{+0.004(+0.028)}$ \\
                \hline
                $R_{K_1^T(1400)}$& SM & $0.893_{-0.003}^{+0.004}$ & $0.984_{-0.000}^{+0.001}$ & $0.997_{-0.000}^{+0.000}$ \\
                \hline	
                $R_{K_1^T(1400)}$& S1  & $0.947_{-0.006(-0.007)}^{+0.005(+0.008)}$ & $0.945_{-0.047(-0.002)}^{+0.068(+0.002)}$ & $0.782_{-0.002(-0.024)}^{+0.003(+0.026)}$ \\
                $R_{K_1^T(1400)}$& S2 & $0.911_{-0.001(-0.004)}^{+0.001(+0.004)}$ & $0.798_{-0.018(-0.027)}^{+0.031(+0.024)}$ & $0.792_{-0.005(-0.025)}^{+0.003(+0.022)}$ \\
                $R_{K_1^T(1400)}$& S3 & $0.910_{-0.001(-0.012)}^{+0.001(+0.012)}$ & $0.817_{-0.021(-0.087)}^{+0.035(+0.087)}$ & $0.802_{-0.005(-0.079)}^{+0.003(+0.079)}$ \\
                $R_{K_1^T(1400)}$& S4 & $0.890_{-0.004(-0.021)}^{+0.005(+0.021)}$ & $0.753_{-0.007(-0.047)}^{+0.016(+0.047)}$ & $0.822_{-0.009(-0.095)}^{+0.006(+0.095)}$ \\
                $R_{K_1^T(1400)}$& S5 & $0.909_{-0.001(-0.005)}^{+0.001(+0.005)}$ & $0.874_{-0.018(-0.033)}^{+0.029(+0.033)}$ & $0.842_{-0.003(-0.032)}^{+0.002(+0.032)}$ \\
                $R_{K_1^T(1400)}$& S6 & $0.950_{-0.007(-0.008)}^{+0.005(+0.009)}$ & $0.947_{-0.048(-0.003)}^{+0.069(+0.003)}$ & $0.788_{-0.002(-0.026)}^{+0.004(+0.027)}$ \\
                \hline
	\end{tabular}		
	\caption{SM and NP predictions for the LFUV ratios $R_{K_1^{(L,\,T)}(1400)}$, with $\theta_{K_1}=34^{\circ}$, in different bins. The first errors listed are due to the uncertainties of the form factors, and the second errors are due to the $1\sigma$ range of the best-fit Wilson coefficients in different NP scenarios.}\label{tab:RK1140034bin}
	\end{center}
\end{table*}
\begin{table*}[!htbp]
	\begin{center}
    \setlength\tabcolsep{1.3pt}
			\begin{tabular}{|ccccc|}
				\hline
				Observable & Scenario & $q^2$/GeV$^2: [0.045,1]$ & $q^2$/GeV$^2: [1,6]$  & $q^2$/GeV$^2: [14,q^2_{\text{max}}]$ \\
				\hline
				$R_{\Lambda}$& SM & $0.935\pm0.024$ & $1.001\pm0.008$ & $0.999\pm0.000$ \\
                \hline	
                $R_{\Lambda}$& S1  & $0.896\pm0.038_{(-0.004)}^{(+0.004)}$ & $0.838\pm0.024_{(-0.016)}^{(+0.018)}$ & $0.785\pm0.001_{(-0.024)}^{(+0.025)}$ \\
                $R_{\Lambda}$& S2 & $0.879\pm0.030_{(-0.007)}^{(+0.006)}$ & $0.805\pm0.012_{(-0.024)}^{(+0.021)}$ & $0.791\pm0.001_{(-0.025)}^{(+0.022)}$ \\
                $R_{\Lambda}$& S3 & $0.885\pm0.027_{(-0.021)}^{(+0.021)}$ & $0.817\pm0.013_{(-0.076)}^{(+0.076)}$ & $0.801\pm0.001_{(-0.079)}^{(+0.079)}$ \\
                $R_{\Lambda}$& S4 & $0.880\pm0.024_{(-0.019)}^{(+0.019)}$ & $0.813\pm0.011_{(-0.073)}^{(+0.073)}$ & $0.820\pm0.002_{(-0.094)}^{(+0.094)}$ \\
                $R_{\Lambda}$& S5 & $0.899\pm0.027_{(-0.008)}^{(+0.008)}$ & $0.863\pm0.013_{(-0.030)}^{(+0.030)}$ & $0.842\pm0.001_{(-0.032)}^{(+0.032)}$ \\
                $R_{\Lambda}$& S6 & $0.893\pm0.040_{(-0.004)}^{(+0.004)}$ & $0.834\pm0.025_{(-0.018)}^{(+0.019)}$ & $0.768\pm0.001_{(-0.027)}^{(+0.029)}$ \\
                \hline
                $R_{\Lambda^0}$& SM  & $1.013\pm0.088$ & $1.004\pm0.010$ & $1.000\pm0.001$ \\
                \hline
                $R_{\Lambda^0}$& S1  & $0.805\pm0.082_{(-0.023)}^{(+0.024)}$ & $0.790\pm0.010_{(-0.023)}^{(+0.025)}$ & $0.784\pm0.001_{(-0.024)}^{(+0.026)}$ \\
                $R_{\Lambda^0}$& S2 & $0.800\pm0.062_{(-0.026)}^{(+0.023)}$ & $0.792\pm0.006_{(-0.026)}^{(+0.022)}$ & $0.796\pm0.001_{(-0.025)}^{(+0.022)}$ \\
                $R_{\Lambda^0}$& S3 & $0.809\pm0.064_{(-0.082)}^{(+0.082)}$ & $0.802\pm0.006_{(-0.081)}^{(+0.081)}$ & $0.805\pm0.001_{(-0.078)}^{(+0.078)}$ \\
                $R_{\Lambda^0}$& S4 & $0.826\pm0.063_{(-0.096)}^{(+0.096)}$ & $0.819\pm0.007_{(-0.095)}^{(+0.095)}$ & $0.827\pm0.002_{(-0.096)}^{(+0.096)}$ \\
                $R_{\Lambda^0}$& S5 & $0.860\pm0.084_{(-0.034)}^{(+0.034)}$ & $0.845\pm0.008_{(-0.033)}^{(+0.033)}$ & $0.845\pm0.001_{(-0.032)}^{(+0.032)}$ \\
                $R_{\Lambda^0}$& S6 & $0.795\pm0.082_{(-0.025)}^{(+0.026)}$ & $0.784\pm0.011_{(-0.026)}^{(+0.027)}$ & $0.769\pm0.002_{(-0.028)}^{(+0.029)}$ \\
                \hline
                $R_{\Lambda^1}$& SM  & $0.901\pm0.004$ & $0.987\pm0.004$ & $0.998\pm0.000$ \\
                \hline
                $R_{\Lambda^1}$& S1  & $0.936\pm0.007_{(-0.005)}^{(+0.005)}$ & $1.064\pm0.081_{(-0.016)}^{(+0.018)}$ & $0.786\pm0.001_{(-0.023)}^{(+0.025)}$ \\
                $R_{\Lambda^1}$& S2 & $0.914\pm0.001_{(-0.003)}^{(+0.002)}$ & $0.864\pm0.058_{(-0.025)}^{(+0.022)}$ & $0.788\pm0.001_{(-0.025)}^{(+0.022)}$ \\
                $R_{\Lambda^1}$& S3 & $0.913\pm0.001_{(-0.008)}^{(+0.008)}$ & $0.886\pm0.058_{(-0.081)}^{(+0.081)}$ & $0.798\pm0.001_{(-0.080)}^{(+0.080)}$ \\
                $R_{\Lambda^1}$& S4 & $0.900\pm0.004_{(-0.014)}^{(+0.014)}$ & $0.790\pm0.042_{(-0.050)}^{(+0.050)}$ & $0.814\pm0.003_{(-0.094)}^{(+0.094)}$ \\
                $R_{\Lambda^1}$& S5 & $0.912\pm0.001_{(-0.003)}^{(+0.003)}$ & $0.927\pm0.040_{(-0.030)}^{(+0.030)}$ & $0.840\pm0.001_{(-0.033)}^{(+0.033)}$ \\
                $R_{\Lambda^1}$& S6 & $0.938\pm0.008_{(-0.005)}^{(+0.006)}$ & $1.069\pm0.083_{(-0.019)}^{(+0.020)}$ & $0.767\pm0.002_{(-0.027)}^{(+0.029)}$ \\
                \hline
	\end{tabular}		
	\caption{SM and NP predictions for the LFUV ratios $R_{\Lambda^{(0,\,1)}}$ in different bins. The first errors listed are due to the uncertainties of the form factors, and the second errors are due to the $1\sigma$ range of the best-fit Wilson coefficients in different NP scenarios.}\label{tab:RLambdabin}
	\end{center}
\end{table*}
\clearpage
\bibliographystyle{refstyle}
\bibliography{references}
\end{document}